\documentclass[aps,prl,twocolumn,superscriptaddress,showpacs,footnotebib]{revtex4-2}
\usepackage{amsmath}
\usepackage{bm}
\usepackage{color}
\usepackage{fontenc}
\usepackage{graphicx}
\usepackage{epstopdf}
\usepackage{epsfig}
\usepackage{amsfonts}
\usepackage[naturalnames]{hyperref}
\usepackage{mathrsfs}
\usepackage{hypcap}
\usepackage{verbatim}
\usepackage{tabularx}
\usepackage{bbm}
\usepackage{esvect}
\usepackage{subfigure}
\usepackage{slashed}
\usepackage{slashed}

\newcommand{\abs}[1]{\left|#1\right|}

\usepackage[T1]{fontenc} % if needed

\def\bea{\begin{eqnarray}}
\def\eea{\end{eqnarray}}

\def\ba{\begin{array}}
\def\ea{\end{array}}

\hypersetup{colorlinks=true, citecolor=blue, urlcolor=blue, linkcolor=blue}

\usepackage[dvipsnames]{xcolor}
\usepackage[normalem]{ulem} % for `\sout{}`

\begin{document}

\title{Exciton-roton mode in moir\'e fractional Chern insulators}

\author{Xiaoyang Shen}
\affiliation{State Key Laboratory of Low Dimensional Quantum Physics and
    Department of Physics, Tsinghua University, Beijing, 100084, China}

\author{Zijian Zhou}
\affiliation{State Key Laboratory of Low Dimensional Quantum Physics and
    Department of Physics, Tsinghua University, Beijing, 100084, China}

\author{Ruiping Guo}
\affiliation{State Key Laboratory of Low Dimensional Quantum Physics and
    Department of Physics, Tsinghua University, Beijing, 100084, China}
\affiliation{Institute for Advanced Study, Tsinghua University, Beijing 100084, China}

\author{Renqi Wang}
\affiliation{State Key Laboratory of Low Dimensional Quantum Physics and
    Department of Physics, Tsinghua University, Beijing, 100084, China}

\author{Wenhui Duan}
\email{duanw@tsinghua.edu.cn}
\affiliation{State Key Laboratory of Low Dimensional Quantum Physics and
    Department of Physics, Tsinghua University, Beijing, 100084, China}
\affiliation{Institute for Advanced Study, Tsinghua University, Beijing 100084, China}
\affiliation{Frontier Science Center for Quantum Information, Beijing, China}
\author{Chong Wang}
\email{chongwang@mail.tsinghua.edu.cn}
\affiliation{State Key Laboratory of Low Dimensional Quantum Physics and
    Department of Physics, Tsinghua University, Beijing, 100084, China}
\affiliation{Frontier Science Center for Quantum Information, Beijing, China}
\author{Yong Xu}
\email{yongxu@mail.tsinghua.edu.cn}
\affiliation{State Key Laboratory of Low Dimensional Quantum Physics and
    Department of Physics, Tsinghua University, Beijing, 100084, China}
\affiliation{Frontier Science Center for Quantum Information, Beijing, China}

\date{\today}

\begin{abstract}
%Moir\'e fractional Chern insulators (FCIs) realize  physics in zero magnetic field.
Moir\'e fractional Chern insulators (FCIs) are a novel class of quantum matter that realizes fractional quantum Hall (FQH) physics in zero magnetic field and provides a platform for exploring unconventional collective excitations. Here we show that hybridization between the magneto-roton and moir\'e interband excitations gives rise to an exciton-roton mode absent in continuum FQH systems in the long-wavelength limit. Using exact diagonalization and a variational Bethe-Salpeter equation for twisted MoTe$_2$, we demonstrate that this hybridization is controlled by the quantum geometry and yields a mode that combines excitonic optical response with the characteristic FCI roton minimum. The resulting exciton-roton remains low-lying, with excitation energy below the interband transition, and acquires optical activity, leading to a double-peak spectroscopic signature. These results identify optical spectroscopy as a direct probe of collective excitations in moir\'e FCIs.

    %The discovery of fractional Chern insulators (FCI) in moir\'e systems draws considerable inspiration to explore intriguing excitations arising from the novel phase of matter. Here, we report the study of the particle-hole excitation physics in the moir\'e FCI and show that the excitation structure of moir\'e FCI is dramatically different from the ideal fractional quantum Hall systems. Utilizing a combination of exact diagonalization and the Bethe-Salpeter equation, we uncover a low-lying hybridization between interband transitions and intraband transitions dubbed as exciton-roton hybridization, a unique phenomenon naturally driven by strong moir\'e band-mixing effects. We further unveil intriguing features about the hybridized excitations, including the fractionalized topology inheriting from the FCI ground state and the active optical response tunable via twist angle and interaction strength. The work provides crucial insights into the interplay of correlations, topology, and light-matter interactions in moiré FCIs.

\end{abstract}
\maketitle

{\it Introduction.---}Moir\'e superlattices \cite{Cai2023,Park2023,LiTingxin2023PRX,JuLong2024,zeng2023thermodynamic} have recently emerged as a versatile platform for realizing fractional Chern insulators (FCIs), the lattice analogs of fractional quantum Hall (FQH) states in zero magnetic field. Although the ground states of FCIs \cite{2021li,2021Devakul,abouelkomasan2020,wang2024fractional,reddy2023fractional,yu2023fractional,regnault2011fractional,2024jia,PhysRevLett.133.186602,PhysRevLett.132.036501,PhysRevLett.132.096602} have been extensively studied, the nature of their excitations remains elusive, despite a few preliminary studies \cite{Repellin_2014,PhysRevB.109.245125,shenroton,kousa2025theorymagnetorotonbandsmoire,paul2025shininglightcollectivemodes}. In conventional FQH systems, the prototypical low-energy neutral excitation is the magneto-roton \cite{PhysRevB.33.2481}, a dark intra-Landau-level collective mode described by the celebrated Girvin--MacDonald--Platzman (GMP) ansatz. Recent studies indicate that magneto-rotons in FCIs closely resemble their FQH counterparts: they exhibit the characteristic roton minimum, remaining only weakly active optically, making them challenging to detect experimentally \cite{wolf2024intrabandcollectiveexcitationsfractional,GDPhysRevLett.107.116801}.

Although the \textit{intraband} excitation structure of FCIs appears to closely parallel that of the FQH systems \cite{Repellin_2014,shenroton,PhysRevB.87.245425,PhysRevX.5.021004,PhysRevLett.134.046501}, FCIs admit an additional and far richer class of \textit{interband} excitations \cite{anyontrions,xu2025localizedexcitonslandaulevelmixing}. In the FQH systems, the remote bands are simply the higher Landau levels, leading to a relatively restricted family of interband excitations \cite{PhysRev.123.1242,PhysRevB.33.3810,MagnetoplasmonExcitations}. By contrast, lattice systems can host remote bands with a wide range of dispersion, topology, and quantum geometry \cite{wang2024diversemagneticordersquantum,IDTPhysRevX.13.041026,liu2025fractionalcherninsulatorstopological}, greatly enriching possible interband excitations \cite{yu2024fractional,qiu2024quantumgeometryprobedchiral,wu2019topological}. In addition, previous studies have shown that band mixing can significantly destabilize FCI phases, suggesting that interband processes may dominate the low-energy sector \cite{bandmixingAhmed,yu2024fractional,fluctuationdriven_yu}. Moreover, due to the breaking of continuous translational symmetry \cite{PhysRev.123.1242}, long-wavelength interband excitations can hybridize with the magneto-roton, providing a more effective route for transferring optical weight to low-energy excitations. These considerations suggest that interband excitations are an essential part of the FCI excitation spectrum, and that a proper theory of collective excitations in FCIs should treat intraband and interband excitations on an equal footing.

In this Letter, we uncover collective exciton-roton mode in FCIs. Arising from the strong hybridization between intraband magneto-rotons and interband optical transitions, the exciton-roton mode carries robust excitonic optical activity with the roton minimum of the FCIs. By developing a microscopic two-level model, we reveal that this hybridization is fundamentally controlled by the quantum geometry and the moir\'e bandgap. While pure intraband magneto-rotons remain optically dark, the resulting exciton-roton emerges as a low-energy, optically bright excitation, efficiently transferring spectral weight to previously hidden modes. Furthermore, we demonstrate that the exciton-roton modes induce a distinct double-peak structure in optical spectra. The experimental observation of this signature will provide a direct spectroscopic probe into the intricate collective dynamics of fractionalized topological quantum fluids.

\begin{figure}[h]
    \centering
    \includegraphics[width=0.95\linewidth]{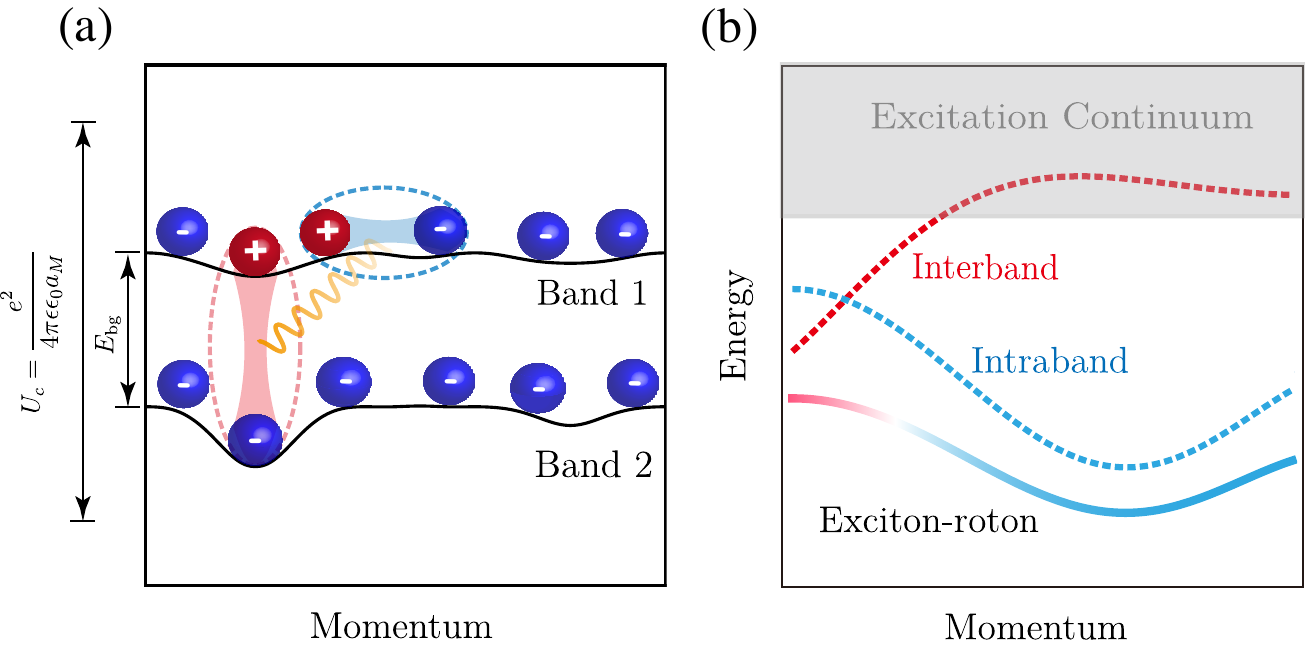}
    \caption{(a) Illustration of the interband transitions and intraband transitions for a moiré fractional Chern insulator. (b) Concept of the exciton-roton mode from hybridization and splitting between collective excitations in the moir\'e FCIs.}
    \label{fig:placeholder}
\end{figure}

% In this Letter, we explore the rich particle-hole excitation physics in moir\'e FCI with a particular focus on the twisted $\text{MoTe}_2$. By restricting to the particle-hole excitations subspace, we find moir\'e band mixing effects induce prominent hybridization between the interband transitions and intraband transitions, which fundamentally change the excitation physics of FCI from ideal FQH. The lowest hybridized bound state constitutes the interband component in the long-wavelength limit, while exhibiting the roton behaviors at finite momentum, and is henceforth dubbed as exciton-roton hybridization. We study the optical response driven by the hybridization effects, and further point out that the three-fold hybridization inherits the intrinsic topology from the ground state and can be characterized by a fractional many-body Chern number.

{\it Two-level model for collective excitations.---}
We begin by demonstrating how interband and intraband excitations inevitably hybridize in moiré fractional Chern insulators, motivating a unified treatment of both excitations on an equal footing.
Consider a moiré FCI ground state labeled as $|\text{GS}\rangle$. The low-energy single-particle excitation physics is spanned by two elementary classes of particle-hole excitations as shown in Fig.~\ref{fig:placeholder}. The first is the intraband sector, $|\phi_{\text{intra}}(\boldsymbol{q})\rangle = C_{v,\boldsymbol{k}+\boldsymbol{q}}^\dagger C_{v,\boldsymbol{k}}|\text{GS}\rangle$, where $C_{v,\boldsymbol{k}}^\dag$ creates a hole in the fractionalized-filled valence band; collective superpositions of such pairs within this sector give rise to the magneto-roton mode \cite{shenroton}. The second is the interband sector, $|\phi_{\text{inter}}(\boldsymbol{q})\rangle= C_{c,\boldsymbol{k}^\prime+\boldsymbol{q}}^\dagger C_{v,\boldsymbol{k}^\prime}|\text{GS}\rangle$, describing optical transitions from the first moir\'e valence band to the conduction band \cite{qiu2024quantumgeometryprobedchiral}. While these two sectors are orthogonal in the non-interacting limit, the Coulomb interaction between different bands explicitly couples them. At $\theta = 3.5^\circ,\epsilon = 5$, the interaction strength scale $U_c = \frac{e^2}{4\pi\epsilon_0\epsilon a_M}\sim 50\text{ meV}$ becomes comparable to the bandgap $E_{\text{bg}} \sim 15\text{ meV}$ (see SM Sec. I \cite{supplemental} for details), this coupling drives a prominent hybridization, dramatically reshaping the excitation structure compared to the ideal FQH limit.
\begin{figure*}[t]
    \centering
    \includegraphics[width=1\linewidth]{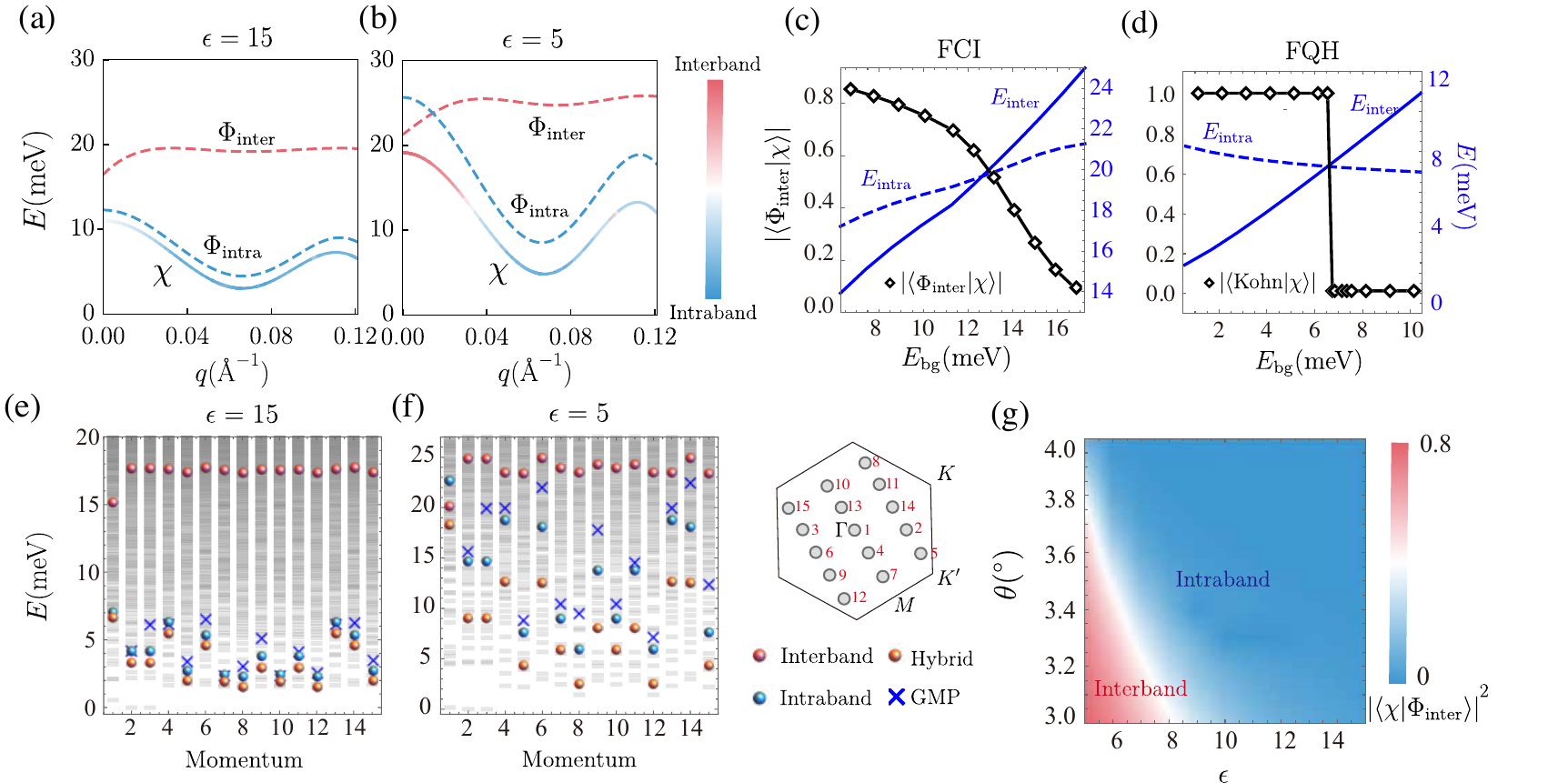}
    \caption{(a,b) The dispersion for the lowest interband transition $\Phi_{\text{inter}}$, intraband transition $\Phi_{\text{intra}}$, and the hybridization $\chi$ at $\epsilon = 15$ and $\epsilon = 5$, respectively. The color of the hybrid mode labels the intraband and interband components. (c,d) The interband/intraband excitation energy (blue solid/dashed) $E_{\text{inter}}/E_{\text{intra}}$ and the overlap $|\langle\Phi_{\text{inter}}|\chi\rangle|$ change with bandgap in two-band FCI and two-band FQH.  (e,f) The lowest excitations in the many-body spectrum at $\epsilon = 5$ and $ \epsilon = 15$. The geometry of ED cluster is given and four excitations are labeled: interband (pink), hybrid (orange), intraband (cyan), GMP (blue cross) (g) The interband overlap $|\langle\chi|\Phi_{\text{inter}}\rangle|^2$ of the $\boldsymbol{q} = 0$ long-wavelength hybrid mode as the function of interaction strength $\epsilon$ and twist angle $\theta$.}
    \label{fig:spectrum}
\end{figure*}

To capture the interplay between the intraband and the interband sectors, we consider a microscopically two-level Hamiltonian between $|\phi_{\text{intra}}(\boldsymbol{q})\rangle$ and $|\phi_{\text{inter}}(\boldsymbol{q})\rangle$. The simplified Hamiltonian yields:
\begin{equation}
    H_{\text{eff}}(\boldsymbol{q}) =
    \begin{pmatrix}
        E^0_{\text{intra}}(\boldsymbol{q}) & \Delta^*(\boldsymbol{q})           \\
        \Delta(\boldsymbol{q})             & E^0_{\text{inter}}(\boldsymbol{q})
    \end{pmatrix},
    \label{eq:Heff}
\end{equation}
where $E^0_{\text{intra/inter}}(\boldsymbol{q})$ are the dispersions of the decoupled modes. The central quantity is the off-diagonal coupling $\Delta(\boldsymbol{q}) = \langle\phi_{\text{inter}}(\boldsymbol{q})|H|\phi_{\text{intra}}(\boldsymbol{q})\rangle$, which mixes the two bare branches. This coupling is generated by the band-mixing part of the Coulomb interaction. We estimate its magnitude at the mean-field level and find that the leading  contribution yields (see SM Sec. IV~\cite{supplemental} for details)
\begin{equation}
    \Delta_{\boldsymbol{k}\boldsymbol{k}^\prime}(\boldsymbol{q}) \propto  \sum_{\boldsymbol{G}} V(\boldsymbol{q}+\boldsymbol{G}) \mathcal{I}_{cv}(\boldsymbol{k}^\prime, \boldsymbol{q}+\boldsymbol{G}) \mathcal{I}_{vv}^*(\boldsymbol{k}, \boldsymbol{q}+\boldsymbol{G}).
    \label{eq:Delta_Exact}
\end{equation}
Here $\mathcal{I}_{nm}(\boldsymbol{k}, \boldsymbol{q}) = \langle u_{n, \boldsymbol{k}+\boldsymbol{q}} | u_{m, \boldsymbol{k}} \rangle$ is the form factor, and the sum runs over reciprocal lattice vectors $\boldsymbol{G}$. The key point is that the same coupling is fixed by quantum geometry: the leading order expansion at small $\boldsymbol{q}$
yields $|\Delta(\boldsymbol{q})|^2 \propto (U_c^2/a_M^2) \sum_{\mu,\nu} \hat{q}_\mu \hat{q}_\nu g_{\mu\nu}$, so the angle-averaged coupling obeys \cite{2021wang,Mera_2021,2014Roy}
\begin{equation}\label{eq:QuantumMetric}
    \langle |\Delta_{\boldsymbol{k}\boldsymbol{k}^\prime}|^2 \rangle_{\theta} \propto \text{tr}g_{\boldsymbol{k}} \geq |\Omega_{\boldsymbol{k}}|,
\end{equation}
where the lower bound follows from the relation between the quantum metric and the Berry curvature. For a topological flat band, this keeps $\Delta$ generically nonzero and produces a robust level repulsion in Eq.~(\ref{eq:Heff}). The resulting splitting between the two hybridized branches scales as $\Delta_{\text{eff}}= \langle\abs{\Delta}^2\rangle_{\theta}/\abs{E^0_{\text{inter}}-E^0_{\text{intra}}}\propto U_c^2/E_{\text{bg}}$, pushing the lower branch downward into the exciton-roton mode.

% we relate the interband term to the non-Abelian Berry connection $\boldsymbol{\mathcal{A}}_{cv}(\boldsymbol{k}) = \langle u_{c, \boldsymbol{k}} | i\nabla_{\boldsymbol{k}} | u_{v, \boldsymbol{k}} \rangle$, while $\mathcal{I}_{vv} \approx 1$. The $1/q$ divergence of the Coulomb potential compensates for the linear vanishing of the geometric term, yielding a finite coupling: $\Delta(\boldsymbol{q} \to 0) \approx i \mathcal{V}_0 \left( \hat{\boldsymbol{q}} \cdot \boldsymbol{\mathcal{A}}_{cv}(\boldsymbol{k}) \right)$, where $\mathcal{V}_0 = 2\pi e^2/(\epsilon A)$ and 

% Crucially, the magnitude of the level repulsion is governed by the Fubini-Study quantum metric $g_{\mu\nu}$ of the valence band. The squared coupling strength relates to the metric tensor as: $|\Delta(\boldsymbol{q})|^2 \approx \mathcal{V}_0^2 \sum_{\mu,\nu} \hat{q}_\mu \hat{q}_\nu g_{\mu\nu}(\boldsymbol{k})$, where $g_{\mu\nu}(\boldsymbol{k}) \approx \text{Re}(\mathcal{A}_{cv}^\mu \mathcal{A}_{cv}^{*\nu})$ within two band system.

For a general collective excitation expressed as a coherent superposition of particle-hole pairs, $|\Psi(\boldsymbol{q})\rangle = \sum_{\boldsymbol{k}} \psi_{\boldsymbol{k}} C^\dagger_{\boldsymbol{k}+\boldsymbol{q}} C_{\boldsymbol{k}}|\text{GS}\rangle$, the total coupling strength becomes a coherent modulation of the quantum metric $g_{\mu\nu}(\boldsymbol{k})$ over the Brillouin zone, convolved with the internal envelope functions of the magneto-roton and the interband transitions (see SM Sec. IV \cite{supplemental} for details).

The microscopic model reveals that the hybridization is governed by two factors: the band mixing parameters $\kappa = U_c/E_{\text{bg}}$ and the quantum metric dictating the geometric matrix element. This implies that the particle-hole excitations in moir\'e FCIs are highly tunable by the twisted angle and interaction strength, allowing the system to be switched between weak perturbative mixing and strong resonant hybridization.
While this effective model provides a physically transparent picture, the higher-order connected correlations are not included, which are crucial in tMoTe$_2$ FCIs. To rigorously explore the strong-coupling regime and capture the many-body effects, we deploy exact numerical methods in the following section.

    {\it ED-variational method and the exciton-roton mode.---} To resolve the excitation spectrum microscopically and identify the origin of the hybridization, we employ a sector-selective variational approach built on the exact FCI ground state. For momentum $\boldsymbol{q}$, a collective excitation is written as a coherent superposition of particle-hole pairs,
$|\Psi_\lambda(\boldsymbol{q})\rangle = \sum_{\boldsymbol{k},\alpha} \psi_{\boldsymbol{k},\alpha}^\lambda(\boldsymbol{q}) \hat{O}_{\boldsymbol{k},\alpha}^\dagger(\boldsymbol{q}) |\text{GS}\rangle$,
where $\alpha \in \{\text{intra}, \text{inter}\}$ labels the intraband and interband sectors. Explicitly,
$\hat{O}_{\boldsymbol{k},\text{intra}}^\dagger(\boldsymbol{q}) = C_{v,\boldsymbol{k}+\boldsymbol{q}}^\dagger C_{v,\boldsymbol{k}}$
and
$\hat{O}_{\boldsymbol{k},\text{inter}}^\dagger(\boldsymbol{q}) = C_{c,\boldsymbol{k}+\boldsymbol{q}}^\dagger C_{v,\boldsymbol{k}}$.
Minimizing the excitation energy within this variational manifold leads to the Bethe-Salpeter equation (BSE):

\begin{equation}\label{eq:BSE_Generalized}\sum_{\boldsymbol{k}', \beta} \Big[ \mathcal{H}_{\boldsymbol{k}\boldsymbol{k}'}^{\alpha\beta}(\boldsymbol{q}) - \varepsilon_\lambda(\boldsymbol{q}) \mathcal{S}_{\boldsymbol{k}\boldsymbol{k}'}^{\alpha\beta}(\boldsymbol{q}) \Big] \psi_{\boldsymbol{k}',\beta}^\lambda = 0.\end{equation}

Here, $\varepsilon_\lambda(\boldsymbol{q})$ is the excitation energy of the $\lambda$-th mode, $E_{\text{GS}}$ is the ground-state energy, and $\mathcal{H}^{\alpha\beta}$ is the projected kernel of the microscopic many-body Hamiltonian $H$,
$\mathcal{H}_{\boldsymbol{k}\boldsymbol{k}'}^{\alpha\beta}(\boldsymbol{q}) = \langle \text{GS} | \hat{O}_{\boldsymbol{k},\alpha}(\boldsymbol{q}) (H - E_{\text{GS}}) \hat{O}_{\boldsymbol{k}',\beta}^\dagger(\boldsymbol{q}) | \text{GS} \rangle$,
and $\mathcal{S}^{\alpha\beta}$ is the overlap matrix,
$\mathcal{S}_{\boldsymbol{k}\boldsymbol{k}'}^{\alpha\beta}(\boldsymbol{q}) = \langle \text{GS} | \hat{O}_{\boldsymbol{k},\alpha}(\boldsymbol{q}) \hat{O}_{\boldsymbol{k}',\beta}^\dagger(\boldsymbol{q}) | \text{GS} \rangle$.
Both matrices are evaluated in the same two-band microscopic model for twisted MoTe$_2$ used in the exact-diagonalization calculation; details of the Hamiltonian and numerical setup are given in the SM \cite{supplemental}.

% \begin{equation}
% \begin{aligned}
%    |\Phi(\boldsymbol{Q})\rangle  & = \sum_{n}f_{n}(\boldsymbol{Q})|n,\boldsymbol{Q}\rangle = \sum_{i}\sum_{\boldsymbol{k}}g_{i,\boldsymbol{k}}(\boldsymbol{Q})|i,\boldsymbol{k},\boldsymbol{Q}\rangle,\\ &g_{i\boldsymbol{k}}(\boldsymbol{Q}) = \sum_{n}f_{n}(\boldsymbol{Q})u^{n}_{i\boldsymbol{k}}(\boldsymbol{Q})
% \end{aligned}
%     \end{equation}

% \begin{equation}
% \begin{aligned}
% \label{BSE}
%     \varepsilon_m(\boldsymbol{q}) c_{m}=\sum_{m^\prime} h_{m,m^\prime}(\boldsymbol{q}) c_{m^\prime}^{},
% \\
%     h_{m^{\prime}, m}(\boldsymbol{q})=\langle  m| H(\boldsymbol{q})\left| m^{\prime}\right\rangle.
%     \end{aligned}
% \end{equation}

Equation~\ref{eq:BSE_Generalized} provides a general framework for collective excitations in moir\'e FCIs. The coefficients $\psi_{\boldsymbol{k},\alpha}^\lambda$ are determined variationally from the BSE, rather than fixed to the GMP form
$\psi_{\boldsymbol{k}}^\text{GMP} = \langle u_{v,\boldsymbol{k+q}}|u_{v,\boldsymbol{k}}\rangle$.
This is important in moir\'e FCIs for two reasons: it incorporates band-mixing effects and lattice effects through moir\'e Hamiltonian $H$, and it remains applicable in the long-wavelength limit $\boldsymbol{q}\to 0$, where the GMP ansatz trivially reduces to the ground state \cite{PhysRevB.33.2481}. The same formalism also lets us isolate the origin of the hybridization by restricting the summation indices in Eq.~(\ref{eq:BSE_Generalized}). We denote the lowest eigenstate obtained in the intraband-only, interband-only, and full two-band variational spaces as the bare magneto-roton $|\Phi_{\text{intra}}\rangle$, the bare interband transition $|\Phi_{\text{inter}}\rangle$, and the hybrid exciton-roton mode $|\chi\rangle$, respectively.

% It is worth pointing out that the ED-BSE method has several advantages when studying the excitations in moir\'e FCI. First, the basis is momentum-resolved and is capable of capturing the momentum dependence of the excitations. Secondly, it naturally incorporates the moir\'e band mixing by including the adjacent bands. The intraband excitations, interband excitations, and the intraband-interband hybridization can be studied by focusing on different numbers of bands. The intraband excitations are generally characterized to be $ \sum_{\boldsymbol{k}}f_{\boldsymbol{k}}(\boldsymbol{q})C_{v, \boldsymbol{k}+\boldsymbol{q}}^{\dagger}C_{v,\boldsymbol{k}}\left|\mathrm{GS}\right\rangle$ when the pairing is fully intraband, $f_{\boldsymbol{k}}$ is the coefficients determined from solving Eq.~\ref{BSE}. The celebrated GMP ansatz $ \hat{\bar{\rho}}(\boldsymbol{q})|\text{GS}\rangle = \sum_{\boldsymbol{k}}\langle u_{v,\boldsymbol{k}+\boldsymbol{q}}|u_{v,\boldsymbol{k}}\rangle C_{v,\boldsymbol{k+q}}^\dagger C_{v,\boldsymbol{k}}|\text{GS}\rangle$\cite{PhysRevB.33.2481,shenroton}, used to characterize magneto-rotons in FQH is a subset of the intraband basis with fixed coefficients $f_{i\boldsymbol{k}}(\boldsymbol{q}) = \langle u_{v,\boldsymbol{k}+\boldsymbol{q}}|u_{v,\boldsymbol{k}}\rangle$. Besides, while the GMP ansatz trivially reduces to the ground state at $\boldsymbol{q} = 0$, ED-BSE method is capable of directly capturing the long-wavelength excitations, which are crucial for the optical property in the moir\'e FCI. 

We first consider the weak band-mixing regime at $\theta = 3.5^\circ$ and $\epsilon = 15$. The long-wavelength interband excitation gap,
$E_{\text{inter}} = \varepsilon_\text{inter}(0) \approx 15.04\text{ meV}$,
lies only slightly above the noninteracting bandgap $E_{\text{bg}} \approx 12.32\text{ meV}$. This indicates that, at weak interaction strength, the interband transition remains close to the FQH Kohn mode \cite{PhysRev.123.1242}. At the same time, the intraband branch is well separated in energy and retains the magneto-roton character; see Fig.~\ref{fig:spectrum}(a)(e). Accordingly, the hybridization is weak. The lowest hybrid mode nearly coincides with the intraband branch and contains only a small interband admixture,
$|\langle\Phi_{\text{inter}}|\chi\rangle|\approx 4.56\%$,
as shown in Fig.~\ref{fig:spectrum}(g).

% However, this does not mean moir\'e band mixing effects on excitations can be neglected. As illustrated in the many-body spectrum Fig.~\ref{fig:spectrum}(d), the hybrid mode (orange) achieves a lower energy than the pure intraband transition (red) at finite momentum. Near the roton minimum (e.g., momentum sector \{5,8,12,15\}), the hybridization from our calculation precisely constitutes the lowest neutral excitations in the many-body spectrum. Especially, at roton minima, $E_{\chi} = 2.64\text{ meV} < E_{\text{intra}} = 3.32\text{ meV}$, with a discrepancy $0.68\text{ meV}$ comparable to the neutral bandgap $\Delta_{\text{neutral}} \approx 1.54\text{ meV}$. This suggests the inclusion of band mixing effects yields the weak hybrid mode, which constitutes the lowest neutral excitation, rather than a pure intraband branch. 

% To distinguish from the FQH, we further calculate the energy of the GMP mode (blue cross), which is the lowest neutral excitations in the strong field limit of FQH. The GMP ansatz energy is $1.5\text{ meV}$ higher than the hybrid mode near the roton minima (e.g., momentum sector \{5,8,12,15\}), a disparity even exceeding the neutral gap $\Delta_{\text{neutral}}$. This suggests the breakdown of the single-mode approximation due to band-mixing effects in the FCI.

We next turn to the strong band-mixing regime, where the hybrid excitation becomes nontrivial. At $\theta = 3.5^\circ$ and $\epsilon = 5$, the interband and intraband branches are comparable in energy and cross near $q\approx 0.02\text{ \AA}^{-1}$, as shown in Fig.~\ref{fig:spectrum}(b). Their coupling therefore becomes strong already in the long-wavelength limit. Relative to the bare interband transition, the hybrid mode is pushed downward by
$\Delta = E_{\text{inter}}-E_{\chi}\approx 1.86\text{ meV}$,
consistent with the level repulsion expected from the microscopic two-level picture. The resulting branch lies below the interband transition gap, carries substantial excitonic weight near $\boldsymbol{q}=0$, and still develops a roton minimum at finite momentum $q_{\text{min}} \approx 0.06\text{ \AA}^{-1}$, as seen from the dispersion of the $\chi$ mode in Fig.~\ref{fig:spectrum}(b). This mode is therefore both excitonic and roton-like, and we refer to it as the exciton-roton.

\begin{figure}[h]
    \centering
    \includegraphics[width=0.95\linewidth]{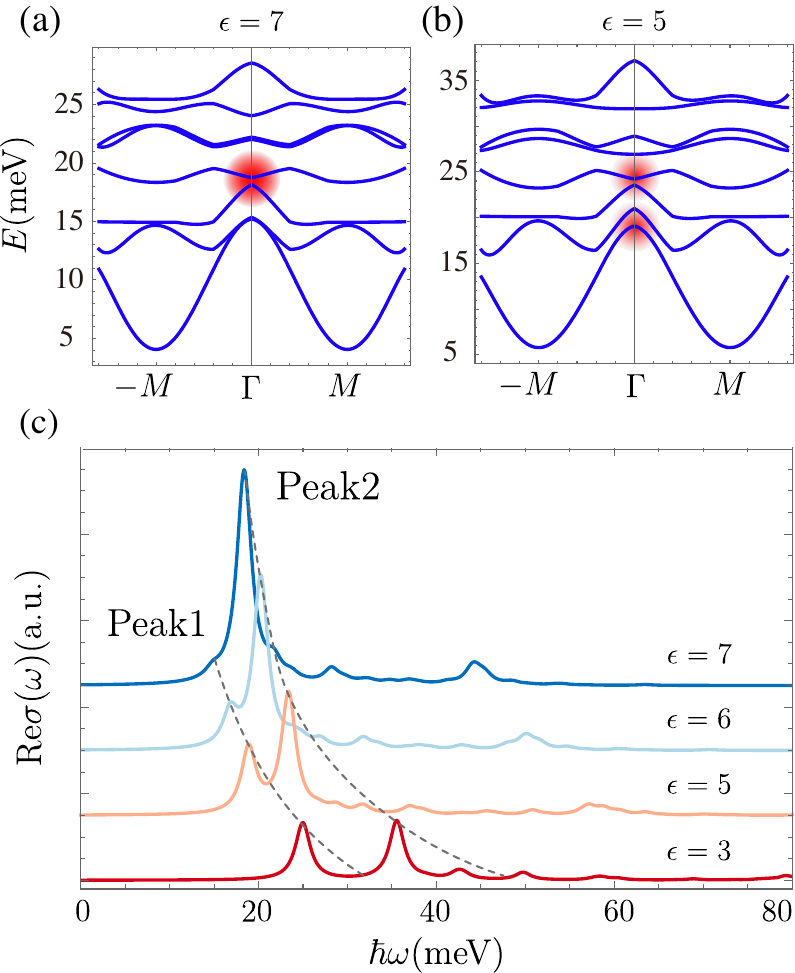}
    \caption{(a,b). The band structure and the optical response of the collective excitations at interaction strengths $\epsilon = 7$ and $\epsilon = 5$. The optical response $\text{Re}\sigma(\omega)$ of the long-wavelength mode is labeled in red. (c) The doublet peaks of the optical response $\text{Re}\sigma(\omega)$ with a series of interaction strengths $\epsilon = 3-7$. The Dirac delta function is simulated by $\delta(x)=\lim _{\gamma \rightarrow 0} \frac{1}{\pi} \frac{\gamma}{\gamma^2+x^2}$ with broadening parameter $\gamma = 0.5\text{ meV}$.}
    \label{fig:X_band_optical}
\end{figure}

The many-body spectrum in this strong-mixing regime, shown in Fig.~\ref{fig:spectrum}(f), is clearly different from its weak-mixing counterpart in Fig.~\ref{fig:spectrum}(e), although both retain the threefold ground-state degeneracy expected for the FCI \cite{yu2023fractional}. Compared with the weak-mixing case, the energies of the GMP mode and the pure intraband mode are substantially enhanced. By contrast, the hybrid branch still accounts for the lowest neutral excitations near the roton minimum, for example in the momentum sectors \{8,12\} of Fig.~\ref{fig:spectrum}(f). This comparison shows that the interband component is indispensable for the low-energy excitation spectrum of moir\'e FCIs.

Finally, we note that several states near the $\Gamma$ point lie below the hybrid mode but are not captured within the present BSE variational space. They likely signal additional low-lying many-body excitations, such as quadrupole modes generated by the many-body stress tensor operator \cite{haldane2011selfdualitylongwavelengthbehaviorlandaulevel,CGFQHLiou,1636-kl65,long2025spectramagnetorotonchiralgraviton}. We leave a detailed study of these modes to future work \footnote{Note that hybridization of the long-wavelength graviton mode with cavity photons via spatially-varying external fields has been recently studied in continuum FQH systems \cite{PhysRevX.15.021027}. That mechanism operates in the spin-2 sector and requires external symmetry-breaking, distinct from the spin-1 particle-hole hybridization driven by Coulomb interaction considered here.}.
To clarify that the long-wavelength mixing is specific to the lattice FCI, we perform comparative ED-BSE calculations for a FQH system with periodic boundary conditions \cite{2021wang,Shi_2024}. The FQH model retains the lowest Landau level (LLL) and the first Landau level (1LL). To facilitate comparison, we set the interaction strength to $\epsilon = 6$ for the FCI and $\epsilon = 15$ for the FQH, and tune the band mixing parameters $\kappa$ by continuously tuning the moir\'e or Landau level bandgap $E_{\text{bg}}$ (see the SM \cite{supplemental} Sec.V for details). We track the long-wavelength excitation energies of the lowest interband and intraband modes, $E_{\text{inter}}$ and $E_{\text{intra}}$, together with the overlap $|\langle \Phi_{\text{inter}} | \chi \rangle|$ between the lowest hybrid mode $|\chi\rangle$ and the pure interband state $|\Phi_{\text{inter}}\rangle$.

As shown in Fig.~\ref{fig:spectrum}(d), the FQH spectrum exhibits a simple level crossing. The interband mode $E_{\text{inter}}$ tracks the bandgap $E_{\text{bg}}$ almost linearly, identifying it as the Kohn mode, whereas the intraband mode $E_{\text{intra}}$ is nearly insensitive to the gap. As a result, the lowest hybrid state $|\chi\rangle$ switches abruptly from the Kohn mode ($E_{\text{inter}} < E_{\text{intra}}$) to a state orthogonal to it ($E_{\text{inter}} > E_{\text{intra}}$). The absence of mixing is enforced by Kohn's theorem \cite{PhysRev.123.1242}: in a Galilean-invariant system, the $\boldsymbol{q} = 0$ center-of-mass motion decouples completely from interaction-driven relative excitations such as the magneto-roton.
In sharp contrast, the FCI results show a smooth evolution rather than a level crossing. As the bandgap is tuned, the overlap $|\langle \Phi_{\text{inter}} | \chi \rangle|$ changes continuously in Fig.~\ref{fig:spectrum}(c), demonstrating robust mixing already at $\boldsymbol{q}=0$. This long-wavelength hybridization is therefore a genuine lattice effect in moiré FCIs, enabled by the breaking of Galilean invariance that forbids such coupling in continuum FQH systems \cite{PhysRevLett.55.2095}.

    {\it Spectroscopic signals of exciton-roton mode.---} The most direct experimental consequence of the exciton-roton is the emergence of a distinct optical signature. To quantify this, we evaluate the longitudinal optical conductivity $\text{Re}\sigma(\omega)$ using the Kubo formula:
\begin{equation}
    \begin{aligned}
         & \text{Re}\sigma(\omega)  = \frac{1}{N}\sum_{n}\frac{1}{\omega}|\langle\chi_n|\hat{v}|\Psi_0\rangle|^2 \delta(\omega-\varepsilon_n+\varepsilon_0).
    \end{aligned}
\end{equation}
where $|\chi_{n}\rangle$ denotes the $n$-th hybrid eigenstate with energy $\varepsilon_n$, and $\hat{v}$ is the velocity operator, see SM Sec.III \cite{supplemental} for details. We focus on single-particle excitations, as many-body modes (e.g., gravitons) have been shown to be optically insensitive \cite{haldane2011selfdualitylongwavelengthbehaviorlandaulevel,shenroton}.

Fig.~\ref{fig:X_band_optical}(c) illustrates the evolution of the optical conductivity. In the weak-mixing regime (e.g., $\epsilon \gtrsim 10$) shown in Fig.~\ref{fig:X_band_optical}(a), the spectrum exhibits a single dominant peak (Peak 2) corresponding to the interband transition continuum, while the lowest hybrid mode (Peak 1) remains optically dark, consistent with its small interband weight $(|\langle\Phi_{\text{inter}}|\chi\rangle|^2\approx 16.04\%$. However, entering the strong-mixing regime ($\epsilon \lesssim 6$) triggers a dramatic spectral reconstruction. The strong hybridization induces a massive level repulsion between the two branches, see Fig.~\ref{fig:X_band_optical}(b)(c). The splitting of the collective excitations is consistent with the two-level picture discussed in the above section. The repulsion pushes down the lower branch (Peak 1), isolating it from the continuum. This isolation creates a discrete, stable bound exciton state within the interband transition gap. Accompanying this formation is a significant transfer of spectral weight, manifesting as a rapid brightening of Peak 1 ($|\langle\Phi_{\text{inter}}|\chi\rangle|^2\approx 65.55\%$) in Fig.~
\ref{fig:X_band_optical}(b). We note that intraband-only theories \cite{paul2025shininglightcollectivemodes,theoryofmagnetorotonw57n-q4xn} predict optical activity through periodic-potential folding alone; our results below show this mechanism is subdominant to interband admixture in the strong-mixing regime.

The emergence of this bright exciton-roton doublet provides a direct experimental fingerprint of collective excitations in FCIs. As demonstrated in Fig.~\ref{fig:X_band_optical}(c), this optical signature exhibits strong tunability with respect to both interaction strength and twist angle, suggesting that moir\'e band engineering can drive the tMoTe$_2$ FCI from a dark, weak-coupling state into a bright, resonant regime. At $\theta = 3.5^\circ$ and $\epsilon = 5$, the two hybridized branches are separated by $\Delta \approx 1.9\,\mathrm{meV}$ and located at $\hbar\omega \approx 15\text{--}20\,\mathrm{meV}$, placing them in the terahertz to far-infrared window ($3\text{--}5\,\mathrm{THz}$) where both time-domain THz spectroscopy and Fourier-transform infrared absorption~\cite{jepsen2011terahertz} routinely achieve sub-meV resolution on two-dimensional moir\'e samples. The lower branch acquires oscillator strength from ${\sim}16\%$ at $\epsilon \approx 7$ to ${\sim}65\%$ at $\epsilon \approx 5$ of the bare interband weight, producing a strongly tunable doublet in reflectance or absorption. Experimentally, this strong-mixing regime can be accessed through moir\'e band engineering: reducing the effective dielectric screening via thinning of the hBN encapsulation or hydrostatic pressure \cite{PhysRevResearch.5.L032022}, or directly narrowing the moir\'e bandgap by increasing the twist angle \cite{mbe1,mbe2,mbe4,shen2024stabilizingfractionalcherninsulators}. The optical setup recently used to detect the $\nu = -1/3$ FCI in tMoTe$_2$~\cite{opticalsignatures} can be directly extended to such samples to resolve the doublet, and the $1.9\,\mathrm{meV}$ splitting comfortably exceeds the thermal broadening at the FCI's operating temperature ($T \lesssim 2\,\mathrm{K}$~\cite{Cai2023,Park2023}), ensuring the doublet remains spectrally resolved. The recent experiments on ``anyon-trion''~\cite{li2026signatures,anyontrions,sensinganyion,ESF,lu2026excitonanyonbindingfractionalchern} also raise the interest of whether an exciton-roton complex can form when the magneto-roton couples to other types of excitons~\cite{sensinganyion,Zhang_2025}, and could thus be detected via photoluminescence (PL).

    {\it Conclusions.---}
We provide a comprehensive study of the single-particle collective excitation physics in moir\'e fractional Chern insulators. Specifically, we unveil a bright exciton-roton mode in moir\'e FCIs that remains both low-lying and optically detectable. We demonstrate that the excitation is a quantum geometry effect. Our theory, especially Eq.~\ref{eq:BSE_Generalized} provides a general framework to study collective excitations in moir\'e FCIs, and can be readily extended to
study the other excitations, such as the chiral gravitons \cite{xavier2025chiralgravitonslattice,long2025spectramagnetorotonchiralgraviton} and moir\'e excitons \cite{lu2026excitonanyonbindingfractionalchern,PhysRevLett.121.067702,PhysRevB.105.235121,Tran_2019} and different types of moir\'e FCIs, such as Jain-sequence \cite{wang2024fractional,PhysRevB.109.245125}, higher Chern number \cite{nkwx-twdl,PhysRevLett.109.186805} and non-abelian FCIs \cite{PhysRevLett.134.076503,43nq-ntqm,PhysRevLett.133.166503,PhysRevLett.134.066601}. Together, these insights highlight the immense potential of moiré correlated phases as a highly tunable platform for excitation manipulation, promising a rich variety of unexplored optical phenomena in future experiments.

    % Finally, we note that our study focuses on single-particle excitations, leaving many-body modes like chiral gravitons \cite{liang2024evidence,shen2024stabilizingfractionalcherninsulators,xavier2025chiralgravitonslattice} unresolved. These likely correspond to the ambiguous sub-gap states observed near the $\Gamma$ point. The ED-BSE formalism, however, can be naturally extended to the many-body excitation subspace to capture these correlations, a direction we reserve for future investigation.

    % We conclude with some remarks about our findings. In the FQHs, when the system is approximated in the LLL limit \cite{PhysRevB.33.3810,PhysRev.123.1242}, the interband transitions are dominated by the Kohn mode, which solely encodes the collective motion of center mass and is fully decoupled from the interaction. Henceforth, we expect that in the long-wavelength limit, the cyclotron mode will be decoupled from the magneto-roton and not inherit the topology of FQH. 

    % Away from the LLL limit and when the Kohn theorem is inapplicable, we expect the interaction to enter the motion of the center of mass, and the energy of the interband transitions is normalized by the many-body interaction. A celebrated example is the magneto-excitons in the monolayer or bilayer van-de Waals materials under a strong magnetic field \cite{PhysRevLett.121.136804,PhysRevB.108.085438,li2024strongly,katsch2020theory}. The magneto-excitons, however, do not hybridize with the intraband neutral excitations, since the excitation energy is significantly larger than the neutral gap \cite{katsch2020theory}. 

    {\it Acknowledgments.---} We thank Chonghao Wang and Xiaodong Hu for helpful discussion. This work was supported by the National Key Basic Research and Development Program of China (Grant No. 2024YFA1409100), the Fundamental and Interdisciplinary Disciplines Breakthrough Plan of the Ministry of Education of China (Grant No. JYB2025XDXM408), 
    the Basic Science Center Project of NSFC (Grant No. 52388201), the National Natural Science Foundation of China (Grants No. 12334003, No. 12421004 and No. 12361141826), the National Key Basic Research and Development Program of China (Grant No. 2023YFA1406400), Beijing Key Laboratory of Quantum AI, and the Beijing Advanced Innovation Center for Future Chip (ICFC). The calculations were performed at National Supercomputer Center in Tianjin using the Tianhe new generation supercomputer.

\bibliography{reference.bib}

\newpage
\clearpage
\onecolumngrid
\vspace{1cm}
\begin{center}
    {\bf\large Supplemental Materials}
\end{center}
\tableofcontents
\section{Continuum Model and interaction}
The single-valley continuum Hamiltonian of the twisted $\mathrm{MoTe_2}$ \cite{wu2019topological} is
\begin{equation}
    \begin{aligned}
        \left.H_0=\left(\begin{array}{cc}-\frac{\hbar^2(\boldsymbol{k}-\boldsymbol{\kappa}_+)^2}{2m^*}+\Delta_\mathfrak{b}(\boldsymbol{r})&\Delta_T(\boldsymbol{r})\\\Delta_T^\dagger(\boldsymbol{r})&-\frac{\hbar^2(\boldsymbol{k}-\boldsymbol{\kappa}_-)^2}{2m^*}+\Delta_\mathfrak{t}(\boldsymbol{r})\end{array}\right.\right),
    \end{aligned}
\end{equation}
The intralayer moir\'e potential $\Delta_{{t},{b}}$ and the interlayer tunneling term $\Delta_T$ are,
\begin{equation}
    \begin{aligned}
    \label{eqn::continuum}
        \Delta_{{t},{b}}(\boldsymbol{r})=2V\sum_{j=1,3,5}\cos\left(\boldsymbol{G}_j\cdot\boldsymbol{r}+\ell\psi\right), \\\Delta_T(\boldsymbol{r})=w\left(1+e^{-i\boldsymbol{G}_2\cdot\boldsymbol{r}}+e^{-i\boldsymbol{G}_3\cdot\boldsymbol{r}}\right).
    \end{aligned}
\end{equation}
$l=\pm 1$ is the layer index and $\boldsymbol{G}_j,j=1,2\dots,6$ are the moir\'e reciprocal lattice vectors. In our calculations, we take the continuum model parameters from fitting first-principles calculation \cite{wang2024fractional} as $V=20.8\,\mathrm{meV}$, $\psi=107.7{}^\circ$, and $w=-23.8\,\mathrm{meV}$. The monolayer lattice constant $a_0$ of MoTe$_2$ is used to be 3.52 \AA, and the effective mass is used to be $0.62m_e$, $m_e$ is the mass of an electron.

The interacting model of the twisted MoTe$_2$ is:
\begin{equation}
\label{Ham}
    H=H_0+\frac{1}{2 A} :\sum_{\boldsymbol{q}} V(\boldsymbol{q}) \hat{\bar{\rho}}(\boldsymbol{q}) \hat{\bar{\rho}}(-\boldsymbol{q}):
\end{equation}
$H_0$ is the Bistritzer-MacDonald (BM)
Hamiltonian in Eq.~\ref{eqn::continuum}, $A$ denotes the area of the two-dimensional system, $\boldsymbol{q}$ is the transferred momentum, $V(\boldsymbol{q})  =\frac{e^2\tanh (|\boldsymbol{q}| d)}{2\epsilon\epsilon_0 |\boldsymbol{q}|}$ is the dual-gate screened Coulomb interaction, 
$d$ is the gate-to-sample distance, $e$ is the charge of electron, $\epsilon_0,\epsilon$ is the vacuum dielectric constant and relative dielectric constant, $:$ means normal order. $\hat{\bar{\rho}}(\boldsymbol{q}) = \sum_{\eta,\eta^\prime}\hat{\rho}_{\eta\eta^\prime}(\boldsymbol{q}) = \sum_{\eta\eta^\prime}\sum_{\boldsymbol{k}}\langle u_{\eta,\boldsymbol{k}+\boldsymbol{q}}|u_{\eta^\prime,\boldsymbol{k}}\rangle C_{\eta,\boldsymbol{k+q}}^\dagger C_{\eta^\prime ,\boldsymbol{k}},\eta,\eta^\prime = 1,\cdots n$ is the density operator projected to the top $n$ valence bands, $|u_{\eta,\boldsymbol{k}}\rangle$ is the periodic part of the Bloch wavefunction of band $\eta$. In particular, we focus on the 2-band twisted $\text{MoTe}_2$
at filling $\nu = -\frac{1}{3}$, $d = 200\text{ \AA}$, twist angle $\theta = 3^\circ\sim 4^\circ$, and $\epsilon = 3-15$ in the main text. We further define the Coulomb interaction scale $U_c = \frac{e^2}{4\pi \epsilon\epsilon_0a_M}\sim 50\text{ meV}$, $a_M = a_0/\theta\approx 57\text{ \AA}$ is the moir\'e length and the moir\'e bandgap $E_{\text{bg}} = \text{min}E_{1}-\text{max} E_2\sim 15\text{ meV}$, $E_{\eta}$ is the band energy of $\eta$-th band.

\section{Numerical details and finite-size effects of the exact diagonalization}

We carry out the full momentum-resolved ED-BSE calculations of 
twisted MoTe$_2$ in three different variational subspaces: 
(1) the interband transition $|\Phi_\text{inter}(\boldsymbol{q})\rangle$ 
defined in $\{C^\dagger_{c,\boldsymbol{k+q}}C_{v,\boldsymbol{k}}|\text{GS}\rangle\}$, 
(2) the intraband transition $|\Phi_{\text{intra}}(\boldsymbol{q})\rangle$ 
defined in $\{C^\dagger_{v,\boldsymbol{k+q}}C_{v,\boldsymbol{k}}|\text{GS}\rangle\}$, 
and (3) their full hybridization $|\chi(\boldsymbol{q})\rangle$ in the 
two-band subspace 
$\{C^\dagger_{\eta,\boldsymbol{k+q}}C_{\eta',\boldsymbol{k}}|\text{GS}\rangle, 
\eta\in\{c,v\}, \eta'=v\}$, with $\boldsymbol{k}$ running over the first BZ. 
All calculations are performed on momentum-space clusters with periodic 
boundary conditions, using the full two-band ground state obtained from 
exact diagonalization. The largest system size we treat is $N = 18$ 
(e.g., $3 \times 6$).

"At strict $\boldsymbol{q} = 0$, the intraband operator $C^\dag_{v,\boldsymbol{k}}C_{v,\boldsymbol{k}}$ reduces to 
the density operator. The corresponding null mode $|\Phi_0\rangle \propto \sum_{\boldsymbol{k}}C^\dagger_{v,\boldsymbol{k}}C_{v,\boldsymbol{k}}|\text{GS}\rangle $ 
is the ground state itself and 
must be removed from the variational manifold. We diagonalize the 
overlap matrix $S^{\alpha\beta}_{\boldsymbol{kk}^\prime}(\boldsymbol{q}=0) $ and project out the eigenvectors 
with eigenvalue below a threshold, then
solve the BSE in the resulting reduced subspace.

\begin{figure}[h]
    \centering
    \includegraphics[width=0.95\linewidth]{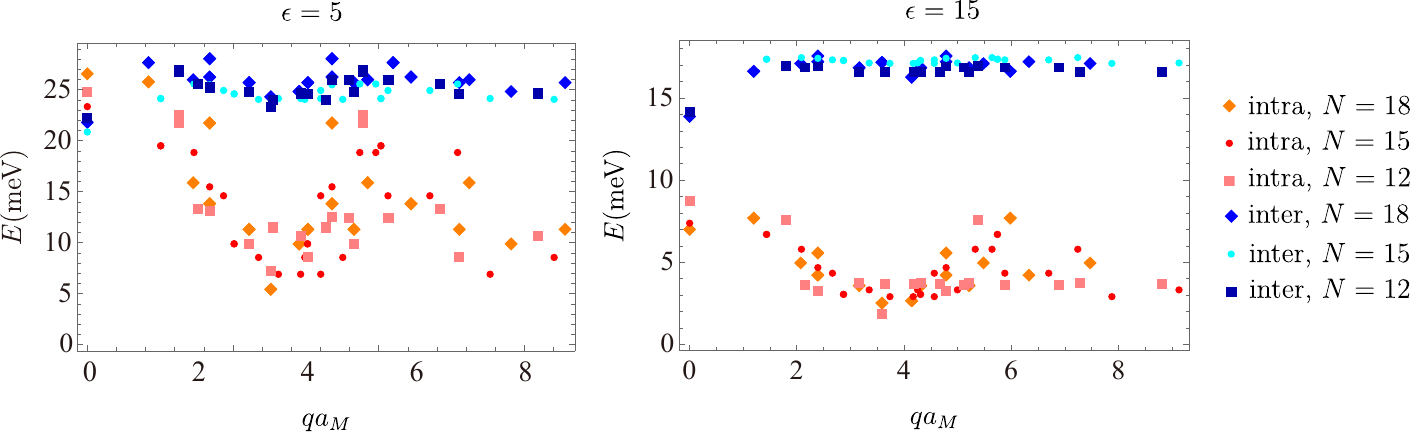}
    \caption{Lowest interband (blue/cyan symbols) and intraband 
    (orange/red/pink symbols) dispersions from ED-BSE on the full 
    two-band ED ground state, for system sizes $N = 12, 15, 18$, 
    at $\epsilon = 5$ (left) and $\epsilon = 15$ (right). Data 
    from different system sizes collapse onto common curves, 
    indicating that finite-size effects are moderate.}
    \label{fig:finite_size}
\end{figure}

We assess finite-size effects by computing the dispersions of the lowest 
interband and intraband modes for cluster sizes $N = 12, 15, 18$ at two 
representative interaction strengths, $\epsilon = 5$ (strong mixing) 
and $\epsilon = 15$ (weak mixing). The results are shown in 
Fig.~\ref{fig:finite_size}. In both interaction regimes, data from 
different system sizes collapse onto common curves within numerical 
uncertainty: the position of the magneto-roton minimum, the energy 
scales of the interband mode, and the long-wavelength behavior of 
both branches are stable against system size. In particular, the 
qualitative features that underlie the main-text conclusions—the 
proximity of $E_\text{inter}$ and $E_\text{intra}$ at strong mixing 
($\epsilon = 5$), and the well-separated branches at weak mixing 
($\epsilon = 15$)—are reproduced consistently across all accessible 
sizes. Finite-size effects are therefore moderate, and the 
exciton-roton phenomenology presented in the main text is not an 
artifact of the finite cluster geometry.

\section{Optical conductivity in many-body system}

According to the Kubo formula, the longitudinal optical conductivity of an interacting many-body system is
\begin{equation}
    \mathrm{Re}\,\sigma_{\rm mb}(\omega) = \frac{\pi e^2}{\omega A}\sum_{m>0}\bigl|\langle\Psi_m|\hat{v}_{\alpha}|\Psi_0\rangle\bigr|^2\,\delta(\omega-E_m+E_0),
    \label{eq:sigma_full}
\end{equation}
where $|\Psi_0\rangle$ and $|\Psi_m\rangle$ denote the exact many-body ground state and excited eigenstates with energies $E_0$ and $E_m$, $A$ is the system area, and $\hat{v}_{\alpha} = \tfrac{i}{\hbar}[H,\hat{r}_\alpha]$ ($\alpha = x,y$) is the $\alpha$-component of the many-body velocity operator. Throughout this section, we suppress the Cartesian index $\alpha$ when no confusion arises.

To make Eq.~\eqref{eq:sigma_full} tractable within the two-band variational framework adopted in the main text, we project $\hat{v}$ onto the low-energy subspace $\mathbb{L}$ spanned by the top valence band ($i=v$) and the lowest conduction band ($i=c$):
\begin{equation}
    \hat{\bar{v}} = \mathcal{P}_{\mathbb{L}}\,\hat{v}\,\mathcal{P}_{\mathbb{L}},
    \qquad
    \mathcal{P}_{\mathbb{L}} = \sum_{\boldsymbol{k}}\sum_{i\in\{v,c\}}|\boldsymbol{k},i\rangle\langle\boldsymbol{k},i|.
\end{equation}
Expanding $\hat{\bar{v}}$ in the Bloch basis,
\begin{equation}
    \hat{\bar{v}} = \sum_{\boldsymbol{k},\boldsymbol{p}}\sum_{i,j}\langle\boldsymbol{k},i|\hat{\bar{v}}|\boldsymbol{p},j\rangle\,C^{\dagger}_{\boldsymbol{k},i}C_{\boldsymbol{p},j},
\end{equation}
the matrix element is
\begin{equation}
    \langle\boldsymbol{k},i|\hat{\bar{v}}|\boldsymbol{p},j\rangle
    = \langle u_{\boldsymbol{k},i}|\,e^{i\boldsymbol{k}\cdot\boldsymbol{r}}\tfrac{i}{\hbar}[H,\hat{\boldsymbol{r}}]e^{-i\boldsymbol{p}\cdot\boldsymbol{r}}\,|u_{\boldsymbol{p},j}\rangle
    = \delta_{\boldsymbol{k},\boldsymbol{p}}\,\langle u_{\boldsymbol{k},i}|\nabla_{\boldsymbol{k}}\bigl(e^{i\boldsymbol{k}\cdot\boldsymbol{r}}H e^{-i\boldsymbol{k}\cdot\boldsymbol{r}}\bigr)|u_{\boldsymbol{k},j}\rangle,
\end{equation}
where momentum conservation enforces $\boldsymbol{k}=\boldsymbol{p}$. Using the identity
\begin{equation}
    \langle u_{\boldsymbol{k},i}|\nabla_{\boldsymbol{k}}H_{\boldsymbol{k}}|u_{\boldsymbol{k},j}\rangle
    = \nabla_{\boldsymbol{k}}\langle u_{\boldsymbol{k},i}|H_{\boldsymbol{k}}|u_{\boldsymbol{k},j}\rangle
    - \langle u_{\boldsymbol{k},i}|H_{\boldsymbol{k}}|\nabla_{\boldsymbol{k}}u_{\boldsymbol{k},j}\rangle
    - \langle\nabla_{\boldsymbol{k}}u_{\boldsymbol{k},i}|H_{\boldsymbol{k}}|u_{\boldsymbol{k},j}\rangle,
\end{equation}
together with $H_{\boldsymbol{k}}|u_{\boldsymbol{k},i}\rangle = \epsilon_i(\boldsymbol{k})|u_{\boldsymbol{k},i}\rangle$, we obtain
\begin{equation}
    \langle\boldsymbol{k},i|\hat{\bar{v}}|\boldsymbol{p},j\rangle
    = \bigl[\,\nabla_{\boldsymbol{k}}\epsilon_i(\boldsymbol{k})\,\delta_{ij}
    + i\bigl(\epsilon_i(\boldsymbol{k})-\epsilon_j(\boldsymbol{k})\bigr)\boldsymbol{A}_{ij}(\boldsymbol{k})\bigr]\,\delta_{\boldsymbol{k},\boldsymbol{p}},
    \label{eq:v_matrix}
\end{equation}
where $\boldsymbol{A}_{ij}(\boldsymbol{k}) = i\langle u_{\boldsymbol{k},i}|\nabla_{\boldsymbol{k}}u_{\boldsymbol{k},j}\rangle$ is the (non-Abelian) Berry connection. In second-quantized form,
\begin{equation}
    \hat{\bar{v}} = \underbrace{\sum_{\boldsymbol{k},i}\nabla_{\boldsymbol{k}}\epsilon_i(\boldsymbol{k})\,C^{\dagger}_{\boldsymbol{k},i}C_{\boldsymbol{k},i}}_{\hat{\bar{v}}_{\text{intra}}\;:\;\text{group-velocity term}}
    \;+\;
    \underbrace{\sum_{\boldsymbol{k}}\sum_{i\neq j}i\bigl(\epsilon_i(\boldsymbol{k})-\epsilon_j(\boldsymbol{k})\bigr)\boldsymbol{A}_{ij}(\boldsymbol{k})\,C^{\dagger}_{\boldsymbol{k},i}C_{\boldsymbol{k},j}}_{\hat{\bar{v}}_{\text{inter}}\;:\;\text{interband Berry-connection term}}.
    \label{eq:v_decomp}
\end{equation}
The two terms have transparent physical meaning: $\hat{\bar{v}}_{\text{intra}}$ encodes the group-velocity contribution within each band, while $\hat{\bar{v}}_{\text{inter}}$ encodes the dipole-allowed interband transitions, which scale with the interband Berry connection $\boldsymbol{A}_{cv}$.

\begin{figure}
    \centering
    \includegraphics[width=0.6\linewidth]{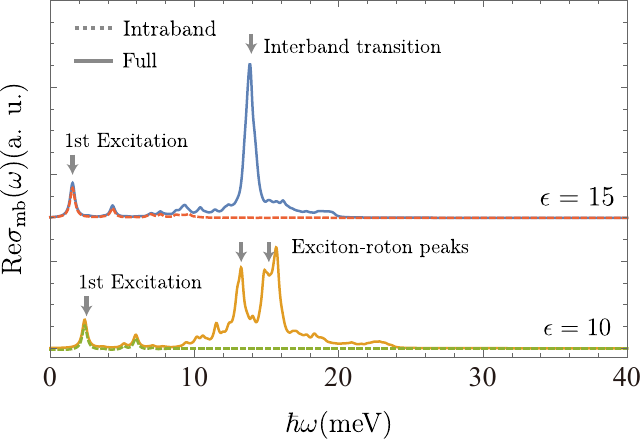}
    \caption{Full many-body optical conductivity 
    $\mathrm{Re}\,\sigma_{\rm mb}(\omega)$ (solid curves) and the 
    intraband-only contribution $\mathrm{Re}\,\sigma_{\rm intra}(\omega)$ 
    (dashed curves) of twisted MoTe$_2$ at twist angle 
    $\theta = 3.5^\circ$ and filling $\nu = -1/3$, for $\epsilon = 15$ (top) 
    and $\epsilon = 10$ (bottom). The lowest many-body excitation 
    is labeled as ``1st Excitation''. The Dirac delta function is 
    broadened by a Lorentzian with $\gamma = 0.15$~meV. Curves are 
    vertically offset for clarity.}
    \label{fig:mb_optical}
\end{figure}

To assess the validity of the variational treatment in the main text 
and to clarify the role of low-lying many-body excitations outside the 
single particle--hole subspace, we compute the full many-body optical 
conductivity $\mathrm{Re}\,\sigma_{\rm mb}(\omega)$ [Eq.~\eqref{eq:sigma_full}] 
by summing over all eigenstates obtained from exact diagonalization 
of the two-band Hamiltonian. We compare it with the intraband-only 
contribution $\mathrm{Re}\,\sigma_{\rm intra}(\omega)$, defined by 
restricting $\hat{\bar{v}}$ to its group-velocity component 
$\hat{\bar{v}}_{\rm intra}$ in Eq.~\eqref{eq:v_decomp}. Results at 
two representative interaction strengths $\epsilon = 15$
and $\epsilon = 10$ are shown in Fig.~\ref{fig:mb_optical}.

At both interaction strengths, the lowest many-body excitation 
(labeled ``1st Excitation'' in Fig.~\ref{fig:mb_optical}) lies well 
below the dominant interband features and carries small optical 
weight relative to the main peaks. These low-lying states may 
correspond to multi-particle--hole excitations such as quadrupole 
or chiral-graviton modes, which have been argued to be weakly 
optically active~\cite{paul2025shininglightcollectivemodes} and lie 
outside the single particle--hole variational manifold of Eq.~(4) 
of the main text. Their relatively small contribution to 
$\mathrm{Re}\,\sigma_{\rm mb}(\omega)$ suggests that the variational 
optical conductivity $\mathrm{Re}\,\sigma_{\chi}(\omega)$ used in 
the main text captures the dominant optical features of the full 
many-body response, although a more complete characterization of 
these low-lying modes is left to future work.

The intraband-only conductivity $\mathrm{Re}\,\sigma_{\rm intra}(\omega)$ 
(dashed curves) remains substantially smaller than the full 
$\mathrm{Re}\,\sigma_{\rm mb}(\omega)$ across the entire frequency 
window, at both $\epsilon = 15$ and $\epsilon = 10$. This is 
consistent with the analysis of Eq.~\eqref{eq:v_decomp}: for nearly 
flat moir\'e bands, $|\nabla_{\boldsymbol{k}}\epsilon_v|$ is small 
compared with $|\epsilon_c - \epsilon_v|\,|\boldsymbol{A}_{cv}|$, and 
the optical weight of the collective modes is therefore largely 
inherited from the interband Berry-connection channel. This 
observation is consistent with intraband-only 
theories~\cite{theoryofmagnetorotonw57n-q4xn,paul2025shininglightcollectivemodes}, 
which note that periodic-potential folding alone produces only weak 
optical activity in the relevant regime, and supports the picture 
that the bright exciton-roton features reported in the main text 
originate predominantly from interband admixture.

The qualitative spectral structure, such as a single dominant peak at 
$\epsilon = 15$ and a doublet-like structure at $\epsilon = 10$, is 
reproduced in the full many-body calculation, consistent with the 
weak- and strong-mixing regimes identified in the main text. We note 
that the doublet onset in the full ED calculation occurs at 
$\epsilon \approx 10$, while the variational two-level treatment 
reaches a comparable doublet at $\epsilon \approx 5$; this difference 
in the crossover parameter reflects correlations 
beyond the single particle--hole manifold. Absolute peak positions 
are correspondingly renormalized. This makes the experimental observation more accessible than the variational estimate suggests.
\section{Effective Two-Level System}

In this section, we construct an effective model to describe the exciton-roton hybridization. We formulate an effective $2\times 2$ Hamiltonian describing the coupling between the intraband magneto-roton and the interband excitation:
\begin{equation}
    \hat{H}_{\text{eff}}(\boldsymbol{q}) = \begin{pmatrix}
        E_{\text{intra}}(\boldsymbol{q})     & \Delta_{\text{BM}}(\boldsymbol{q}) \\
        \Delta_{\text{BM}}^*(\boldsymbol{q}) & E_{\text{inter}}(\boldsymbol{q})
    \end{pmatrix},
\end{equation}
based on the collective modes $|\Phi_{\text{inter}}(\boldsymbol{q})\rangle = \sum_{\boldsymbol{k}_1}f_{\boldsymbol{k}_1}C_{c,\boldsymbol{k}_1+\boldsymbol{q}}^\dagger C_{v,\boldsymbol{k}_1}|\text{GS}\rangle$ and $|\Phi_{\text{intra}}(\boldsymbol{q})\rangle=\sum_{\boldsymbol{k}_2}g_{\boldsymbol{k}_2}C_{v,\boldsymbol{k}_2+\boldsymbol{q}}^\dagger C_{v,\boldsymbol{k}_2}|\text{GS}\rangle$. The diagonal terms are defined as $E(\boldsymbol{q}) = \langle\Phi(\boldsymbol{q})|H|\Phi(\boldsymbol{q})\rangle/\langle\Phi(\boldsymbol{q})|\Phi(\boldsymbol{q})\rangle - E_0$, where $E_0$ is the ground state energy. The off-diagonal coupling term, which drives the resonance between these two modes, is given by:
\begin{equation}
    \Delta_{\text{BM}}(\boldsymbol{q}) = \langle \Phi_{\text{inter}}(\boldsymbol{q})|H|\Phi_{\text{intra}}(\boldsymbol{q})\rangle.
\end{equation}
The coupling is governed by the band-mixing component of the Coulomb interaction:
\begin{equation}
    \label{eq:H_BM}
    H_{\text{BM}} = \frac{1}{A} \sum_{\boldsymbol{Q}, \boldsymbol{k}, \boldsymbol{p}} \mathcal{V}_{\boldsymbol{Q}}(\boldsymbol{k}, \boldsymbol{p}) C_{c, \boldsymbol{k}+\boldsymbol{Q}}^\dagger C_{v, \boldsymbol{p}-\boldsymbol{Q}}^\dagger C_{v, \boldsymbol{p}} C_{v, \boldsymbol{k}} + \text{h.c.},
\end{equation}
where the interaction vertex is given by $\mathcal{V}_{\boldsymbol{Q}}(\boldsymbol{k}, \boldsymbol{p}) = V(\boldsymbol{Q}) \mathcal{I}_{cv}(\boldsymbol{k}, \boldsymbol{Q}) \mathcal{I}_{vv}(\boldsymbol{p}, -\boldsymbol{Q})$.
We evaluate the off-diagonal matrix element $\Delta_{}(\boldsymbol{q}) = \langle \Phi_{\text{inter}}(\boldsymbol{q}) | H_{\text{BM}} | \Phi_{\text{intra}}(\boldsymbol{q}) \rangle$, which involves expectation value of 8 operators. By assuming an empty conduction band, we first perform the contraction of the conduction band operators from $H_\text{BM}$ and $|\Phi_{\text{inter}}\rangle$. This simplifies $\Delta$ to be intraband:
\begin{equation}
    \Delta(\boldsymbol{q}) \approx \langle \mathcal{O}_{\text{valence}}(\boldsymbol{q}) \rangle,\quad \mathcal{O}_{\text{valence}}(\boldsymbol{q}) = \sum_{\boldsymbol{k,p,Q,k_2,k_1}}\mathcal{V}_{\boldsymbol{Q}}(\boldsymbol{k}, \boldsymbol{p}) C_{v, \boldsymbol{k_1}}^\dagger  C_{v, \boldsymbol{p}-\boldsymbol{Q}}^\dagger C_{v, \boldsymbol{p}} C_{v, \boldsymbol{k}} C_{v, \boldsymbol{k}_2+\boldsymbol{q}}^\dagger C_{v, \boldsymbol{k}_2}
\end{equation}

$\boldsymbol{k}_1=\boldsymbol{k-q+Q}$ from the momentum conservation. $\mathcal{O}_{\text{valence}} $ describes an effective intraband interaction that couples the interband transitions and intraband transitions.

To estimate the leading order of $\Delta$, we now apply the mean-field approximation to $\mathcal{O}_{\text{valence}}$ by assuming the band mixing term is not strong enough to destroy the FCI. We further assume the translational invariant charge distribution of the FCI $\langle C_{v, \boldsymbol{k}}^\dagger C_{v, \boldsymbol{k}'} \rangle = \delta_{\boldsymbol{k}, \boldsymbol{k}'} n_{\boldsymbol{k}}, n_{\boldsymbol{k}} \approx \nu$. The six mean-field decomposition channels of the effective interaction are as follows:

\paragraph{(1) The Exchange Channel ($\boldsymbol{Q} \approx \boldsymbol{q}$):}
This channel corresponds to the contraction pairing that links the same momentum transfer in the Coulomb interaction and collective excitations. The contraction is:
\begin{equation}
    \begin{aligned}
        \mathcal{O}_{\text{exc}} & \approx \langle C_{v, \boldsymbol{k}_1}^\dagger C_{v, \boldsymbol{k}} \rangle \langle C_{v, \boldsymbol{p}-\boldsymbol{Q}}^\dagger C_{v, \boldsymbol{k}_2} \rangle \langle C_{v, \boldsymbol{p}} C_{v, \boldsymbol{k}_2+\boldsymbol{q}}^\dagger \rangle
        \approx \delta_{\boldsymbol{Q}, \boldsymbol{q}} \, n_{\boldsymbol{k}_1} n_{\boldsymbol{k}_2} (1 - n_{\boldsymbol{k}_2+\boldsymbol{q}}),                                                                                                                                            \\
        % \Delta_{\text{exc}} &= -\sum_{\boldsymbol{Q}} V(\boldsymbol{Q}) \mathcal{I}_{cv}(\boldsymbol{k}_1, \boldsymbol{Q}) \mathcal{I}_{vv}^*(\boldsymbol{k}_2, \boldsymbol{Q})\delta_{\boldsymbol{Q}, \boldsymbol{q}} \, n_{\boldsymbol{k}_1} n_{\boldsymbol{k}_2} (1 - n_{\boldsymbol{k}_2+\boldsymbol{q}}).
    \end{aligned}
\end{equation}

\paragraph{(2) The Direct Channel ($\boldsymbol{Q} \approx \boldsymbol{k}_1 - \boldsymbol{k}_2$):}
This channel describes the direct scattering between the particle in interband transition and the hole in the intraband transition. The contraction yields:
\begin{equation}
    \mathcal{O}_{\text{dir}} \approx -\langle C_{v, \boldsymbol{k}_1}^\dagger C_{v, \boldsymbol{p}} \rangle \langle C_{v, \boldsymbol{p}-\boldsymbol{Q}}^\dagger C_{v, \boldsymbol{k}_2} \rangle \langle C_{v, \boldsymbol{k}} C_{v, \boldsymbol{k}_2+\boldsymbol{q}}^\dagger \rangle \approx -\delta_{\boldsymbol{Q}, \boldsymbol{k}_1-\boldsymbol{k}_2} \, n_{\boldsymbol{k}_1} n_{\boldsymbol{k}_2} (1 - n_{\boldsymbol{k}_2+\boldsymbol{q}}).
\end{equation}
% \Delta_{\text{dir}} \approx \sum_{\boldsymbol{G}} V(\boldsymbol{k}_{12}+\boldsymbol{G}) \mathcal{I}_{cv}(\boldsymbol{k}_2+\boldsymbol{q}, \boldsymbol{k}_{12}+\boldsymbol{G}) \mathcal{I}_{vv}^*(\boldsymbol{k}_2, \boldsymbol{k}_{12}+\boldsymbol{G}).
Combining these results and restoring the summation over moir\'e reciprocal lattice vectors $\boldsymbol{G}$, we obtain the expression for the $\Delta$ in these two channel:
\begin{equation}
    \Delta_{\text{}}(\boldsymbol{q}) = \frac{\nu^2(1-\nu)}{A} \sum_{\boldsymbol{k}_1, \boldsymbol{k}_2} f_{\boldsymbol{k}_1}^* g_{\boldsymbol{k}_2} \left[ \Delta_{\text{exc}}(\boldsymbol{k}_1, \boldsymbol{k}_2, \boldsymbol{q}) - \Delta_{\text{dir}}(\boldsymbol{k}_1, \boldsymbol{k}_2, \boldsymbol{q}) \right],
\end{equation}
where the explicit momentum-dependent kernels are:
\begin{subequations}
    \begin{align}
        \Delta_{\text{exc}} & = \sum_{\boldsymbol{G}} V(\boldsymbol{q}+\boldsymbol{G}) \mathcal{I}_{cv}(\boldsymbol{k}_1, \boldsymbol{q}+\boldsymbol{G}) \mathcal{I}_{vv}^*(\boldsymbol{k}_2, \boldsymbol{q}+\boldsymbol{G}),                               \\
        \Delta_{\text{dir}} & = \sum_{\boldsymbol{G}} V(\boldsymbol{k}_{12}+\boldsymbol{G}) \mathcal{I}_{cv}(\boldsymbol{k}_2+\boldsymbol{q}, \boldsymbol{k}_{12}+\boldsymbol{G}) \mathcal{I}_{vv}^*(\boldsymbol{k}_2, \boldsymbol{k}_{12}+\boldsymbol{G}).
    \end{align}
\end{subequations}
where we substitue $\boldsymbol{Q} = \boldsymbol{q+G}$, with $\boldsymbol{k}_{12} \equiv \boldsymbol{k}_1 - \boldsymbol{k}_2$.

Due to the orthogonality of the Bloch states, the interband form factor vanishes at exactly $\boldsymbol{q} = 0$. For small but finite $\boldsymbol{q}$, we can expand the form factor up to linear order in momentum:\begin{equation}\label{eq:form_factor_expansion}\mathcal{I}_{cv}(\boldsymbol{k}, \boldsymbol{q}) = \langle u_{c, \boldsymbol{k}+\boldsymbol{q}} | u_{v, \boldsymbol{k}} \rangle \approx \boldsymbol{q} \cdot \langle \nabla_{\boldsymbol{k}} u_{c, \boldsymbol{k}} | u_{v, \boldsymbol{k}} \rangle = -i \boldsymbol{q} \cdot \boldsymbol{A}_{cv}(\boldsymbol{k}),
\end{equation}
where $\boldsymbol{A}_{cv}(\boldsymbol{k}) = i \langle u_{c, \boldsymbol{k}} | \nabla_{\boldsymbol{k}} u_{v, \boldsymbol{k}} \rangle$ is the interband (non-Abelian) Berry connection. Concurrently, the intraband form factor to zeroth order remains $\mathcal{I}_{vv}^*(\boldsymbol{k}_2, \boldsymbol{q}) \approx 1$. Taking account into the Coulomb interaction leads to a finite, macroscopic exchange coupling at $\boldsymbol{q} \to 0$:\begin{equation}\label{eq:Delta_exc_limit}\Delta_{\text{exc}}^{(\boldsymbol{G}=0)}(\boldsymbol{k}_1, \boldsymbol{k}_2, \boldsymbol{q} \to 0) \approx \frac{ e^2}{2\epsilon\epsilon_0} \left( -i \hat{\boldsymbol{q}} \cdot \boldsymbol{A}_{cv}(\boldsymbol{k}_1) \right),\end{equation}where $\hat{\boldsymbol{q}} = \boldsymbol{q}/|\boldsymbol{q}|$ is the unit vector. The Umklapp processes ($\boldsymbol{G} \neq 0$) in the exchange channel do not possess the Coulomb singularity; thus, they scale linearly as $\mathcal{O}(|\boldsymbol{q}|)$ and vanish in the macroscopic limit. The angle average of $\left|\Delta_{\text{exc}}^{(\boldsymbol{G}=0)}\right|^2$ then calculated in the main text is
\begin{equation}
\begin{aligned}
\langle |\Delta|^2 \rangle_{\theta} = &\left[ \frac{\nu^2(1-\nu)}{A} \frac{e^2}{2\epsilon\epsilon_0} \right]^2\langle | \hat{\boldsymbol{q}} \cdot \boldsymbol{A}_{cv} |^2 \rangle_{\theta} \\
=& \left[ \frac{\nu^2(1-\nu)}{A} \frac{e^2}{2\epsilon\epsilon_0} \right]^2\frac{1}{2\pi} \int_0^{2\pi} |A_{cv,x} \cos\theta + A_{cv,y} \sin\theta|^2 d\theta \\
=&\frac{1}{2} \left[ \frac{\nu^2(1-\nu)}{A} \frac{e^2}{2\epsilon\epsilon_0} \right]^2 |\boldsymbol{A}_{cv}(\boldsymbol{k})|^2\\
=& \frac{1}{2} \left[ \frac{\nu^2(1-\nu)}{A} \frac{e^2}{2\epsilon\epsilon_0} \right]^2\operatorname{tr}g(\boldsymbol{k})
\propto [\nu^2(1-\nu)]^2 U_c^2 \left( \frac{\text{tr} g(\boldsymbol{k})}{a_M^2} \right).
\end{aligned}
\end{equation}
Here we have used $g_{ab}(\boldsymbol{k}) \approx \text{Re} [ A_{cv,a} A_{cv,b}^* ]$ in the two-band model and substitue $U_c = \frac{e^2}{4\pi \epsilon\epsilon_0 a_M}$ into the equation.

In stark contrast, the direct channel ($\Delta_{\text{dir}}$) involves the momentum transfer $\boldsymbol{k}_{12} = \boldsymbol{k}_1 - \boldsymbol{k}_2$, which is generally finite and spans the entire moiré Brillouin zone. Consequently, the Coulomb potential $V(\boldsymbol{k}_{12}+\boldsymbol{G})$ is finite. Therefore, in the optical limit, the exciton-roton hybridization is overwhelmingly dominated by the exchange interaction, governed directly by the interband Berry connection:

\begin{equation}\Delta_{\text{}}(\boldsymbol{q} \to 0) \approx -i \frac{ e^2}{2\epsilon\epsilon_0 A} \nu^2(1-\nu) \sum_{\boldsymbol{k}_1, \boldsymbol{k}_2} f_{\boldsymbol{k}_1}^* g_{\boldsymbol{k}_2} \left[ \hat{\boldsymbol{q}} \cdot \boldsymbol{A}_{cv}(\boldsymbol{k}_1) \right].\end{equation}

This result explicitly ties the macroscopic hybridization strength to the topological properties of the system. The magnitude of the mode repulsion is directly bounded by the quantum metric of the bands, as the interband Berry connection is fundamentally related to the off-diagonal components of the quantum geometric tensor. This demonstrates that the exciton-roton splitting is inherently a quantum geometric effect.

The other four terms, as we listed in the following, are subleading:
\paragraph{(3) Diagonal Fock Term :} This term comes from the Fock contraction of the Coulomb interaction
\begin{subequations}
    \begin{align}
        \mathcal{O}_{\text{Fock}} & \approx \langle C_{v, \boldsymbol{k}_1}^\dagger C_{v, \boldsymbol{k}_2} \rangle \langle C_{v, \boldsymbol{p-Q}}^\dagger C_{v, \boldsymbol{k}} \rangle \langle C_{v, \boldsymbol{p}} C_{v,\boldsymbol{k_2+q}}^\dagger \rangle  \approx \delta_{\boldsymbol{k}_1, \boldsymbol{k}_2} \left[ n_{\boldsymbol{k}_1} n_{\boldsymbol{k}_1+\boldsymbol{q}-\boldsymbol{Q}} (1 - n_{\boldsymbol{k}_1+\boldsymbol{q}}) \right], \\
        \Delta_{\text {Fock }}    & =\delta_{\boldsymbol{k}_1, \boldsymbol{k}_2} \sum_{\boldsymbol{Q}} V(\boldsymbol{Q}) \mathcal{I}_{c v}(\boldsymbol{k}, \boldsymbol{Q}) \mathcal{I}_{v v}^*(\boldsymbol{p}, \boldsymbol{Q})\left[n_{\boldsymbol{k}_1} n_{\boldsymbol{k}_1+\boldsymbol{q}-\boldsymbol{Q}}\left(1-n_{\boldsymbol{k}_1+\boldsymbol{q}}\right) \approx 0\right.
    \end{align}
\end{subequations}
In the long-wavelength limit, the interband form factor is proportional to the Berry connection $\sim \boldsymbol{Q}\cdot A_{cv}$, which is odd under inversion, while the intraband form factor and the Coulomb interactions are even $\sim 1$, thus the summation over $\boldsymbol{Q}$ yields approximately zero.
\paragraph{(4) Diagonal Hartree Term:} this term comes from the Hartree contraction of Coulomb interaction
\begin{subequations}
    \begin{align}
        \mathcal{O}_{\text{Hartree}} & \approx \langle C_{v, \boldsymbol{k}_1}^\dagger C_{v, \boldsymbol{k}_2} \rangle \langle C_{v, \boldsymbol{p-Q}}^\dagger C_{v, \boldsymbol{p}} \rangle \langle C_{v, \boldsymbol{k}} C_{v, \boldsymbol{k_2+q}}^\dagger\rangle\approx\delta_{\boldsymbol{k}_1, \boldsymbol{k}_2} \delta_{\boldsymbol{Q},0} \left[ n_{\boldsymbol{k}_1} n_{\boldsymbol{p}} (1 - n_{\boldsymbol{k}_1+\boldsymbol{q}}) \right] \\
        \Delta_{\text{Hartree}}      & = \delta_{\boldsymbol{k}_1, \boldsymbol{k}_2} V(0) \mathcal{I}_{cv}(\boldsymbol{k}_1+\boldsymbol{q}, 0) {\mathcal{I}_{vv}(\boldsymbol{p}, 0)} \left[ n_{\boldsymbol{k}_1} n_{\boldsymbol{p}} (1 - n_{\boldsymbol{k}_1+\boldsymbol{q}}) \right] = 0
    \end{align}
\end{subequations}
which is zero due to $\mathcal{I}_{cv}(\boldsymbol{k}_1+\boldsymbol{q}, 0) = 0$.
\paragraph{(5) Fully Disconnected Term:}
this term comes from contraction
\begin{subequations}
    \begin{align}
        \mathcal{O}_{\text{disc}} & \approx  \langle C^\dagger_{v\boldsymbol{p-Q}}C_{v\boldsymbol{p}}\rangle\langle C^\dagger_{v\boldsymbol{k}_1}C_{v\boldsymbol{k}}\rangle\langle C^\dagger_{v\boldsymbol{k}_2+\boldsymbol{q}}C_{v\boldsymbol{k}_2}\rangle  \approx \delta_{\boldsymbol{q}, 0} \delta_{\boldsymbol{Q}, 0} \cdot \left[ n_{\boldsymbol{k}_1} n_{\boldsymbol{p}} n_{\boldsymbol{k}_2} \right], \\
        \Delta_{\text{disc}}      & = \delta_{\boldsymbol{q}, 0} V(0) {\mathcal{I}_{cv}(\boldsymbol{k}_1, 0)}{\mathcal{I}_{vv}(\boldsymbol{p}, 0)}n_{\boldsymbol{k}_1} n_{\boldsymbol{p}} n_{\boldsymbol{k}_2} = 0
    \end{align}
\end{subequations}
is zero due to $\mathcal{I}_{cv}(\boldsymbol{k}_1, 0) = 0$.

\paragraph{(6) Disconnected Scattering Term:}
Comes from Wick contraction

\begin{subequations}
    \begin{align}
        \mathcal{O}_{\text{disc-scatt}} & \approx -\langle C^\dagger_{v\boldsymbol{k}_1}C_{v\boldsymbol{p}}\rangle\langle C^\dagger_{v\boldsymbol{p-Q}}C_{v\boldsymbol{k}}\rangle\langle C^\dagger_{v\boldsymbol{k}_2+\boldsymbol{q}}C_{v\boldsymbol{k}_2}\rangle \approx -\delta_{\boldsymbol{q}, 0}  \left[ n_{\boldsymbol{k}_1} n_{\boldsymbol{k}_1-\boldsymbol{Q}} n_{\boldsymbol{k}_2} \right] \\
        \Delta_{\text{disc-scatt}}      & =\delta_{\boldsymbol{q}, 0} \sum_{\boldsymbol{Q}} V(\boldsymbol{Q}) \mathcal{I}_{cv}(\boldsymbol{k}_1-\boldsymbol{Q}, \boldsymbol{Q}) \mathcal{I}_{vv}(\boldsymbol{k}_1, -\boldsymbol{Q})n_{\boldsymbol{k}_1} n_{\boldsymbol{k}_1-\boldsymbol{Q}} n_{\boldsymbol{k}_2}\approx 0
    \end{align}
\end{subequations}
The summation in the long-wavelength limit of $\boldsymbol{Q}$ yields zero due to the odd parity.

\section{Hamiltonian for multiband Fractional Quantum Hall system}

In this section, we present the details of the Hamiltonian of the 
multiband FQH system with periodic boundary conditions, following 
\cite{2021wang,Shi_2024}. The setup consists of a torus of area $A$ 
threaded by a uniform magnetic field, with $N_\phi$ flux quanta and 
electrons partially filling the lowest Landau level (LLL) at $\nu = 1/3$. 
The magnetic length is defined as $\ell_B = \sqrt{\hbar/eB}$, and the 
cyclotron energy is $\hbar\omega_c = \hbar eB/m^*$. To facilitate direct 
comparison with the moir\'e FCI calculations in the main text, we 
discretize momenta on a lattice commensurate with the moir\'e Brillouin 
zone, so that
\begin{equation}
    \ell = \sqrt{\frac{\sqrt{3}}{4\pi}}\, a_M, \quad a_M \approx a/\theta,
\end{equation}
plays the role of an effective magnetic length set by the moir\'e 
period $a_M$. We restrict the calculation to a two-Landau-level 
subspace (LLL $+$ 1LL), with the cyclotron gap $\hbar\omega_c$ 
treated as a tunable parameter that controls the strength of inter-LL 
band-mixing effects.

We label the Bloch wavefunction of the $n$-th Landau level as
\begin{equation}
    |\psi^{n\text{LL}}_{\boldsymbol{k}}\rangle \equiv |\boldsymbol{k},n\rangle =  e^{i\boldsymbol{k}\cdot\boldsymbol{r}}|u_{\boldsymbol{k}}^{n\text{LL}}\rangle.
\end{equation}
The interaction matrix element in the Bloch basis reads
\begin{equation}
    \begin{aligned}
        H^{n_1 n_2 n_3 n_4}_{\boldsymbol{k}_1 \boldsymbol{k}_2 \boldsymbol{k}_3 \boldsymbol{k}_4} 
        & \equiv \langle\boldsymbol{k}_1,n_1; \boldsymbol{k}_2,n_2| H_{\text{int}}|\boldsymbol{k}_3,n_3; \boldsymbol{k}_4,n_4\rangle \\
        & = \frac{1}{A} \sum_{\boldsymbol{q}} V_{\boldsymbol{q}}\langle\boldsymbol{k}_1,n_1| e^{i \boldsymbol{q} \cdot \boldsymbol{r}}|\boldsymbol{k}_3,n_3\rangle\langle\boldsymbol{k}_2,n_2| e^{-i \boldsymbol{q} \cdot \boldsymbol{r}}|\boldsymbol{k}_4,n_4\rangle \\
        & = \frac{1}{A} \delta_{\boldsymbol{k}_1+\boldsymbol{k}_2, \boldsymbol{k}_3+\boldsymbol{k}_4} \sum_{\boldsymbol{G}} V_{\boldsymbol{k}_1-\boldsymbol{k}_3-\boldsymbol{G}} \, \mathcal{I}_{\boldsymbol{k}_1,n_1; \boldsymbol{k}_3,n_3}^{-\boldsymbol{G}} \mathcal{I}_{\boldsymbol{k}_2,n_2; \boldsymbol{k}_4,n_4}^{\boldsymbol{k}_3+\boldsymbol{k}_4-\boldsymbol{k}_1-\boldsymbol{k}_2+\boldsymbol{G}},
    \end{aligned}
\end{equation}
where $V_{\boldsymbol{q}} = \frac{e^2\tanh(|\boldsymbol{q}|d)}{2\epsilon_0\epsilon|\boldsymbol{q}|}$ 
is the dual-gate-screened Coulomb interaction with the same parameters 
as those used in the FCI calculations of the main text, and 
$\mathcal{I}^{\boldsymbol{Q}}_{\boldsymbol{k},n;\boldsymbol{p},n^\prime} 
= \langle u_{\boldsymbol{k},n}|e^{i\boldsymbol{Q}\cdot\boldsymbol{r}}|u_{\boldsymbol{p},n^\prime}\rangle$ 
denotes the form factor encoding the overlap between Landau-level Bloch 
states at different momenta and band indices.

In second-quantized form, the total Hamiltonian is $H = H_0 + H_{\text{int}}$, 
with the non-interacting part
\begin{equation}
    H_0 = \sum_{n}\sum_{\boldsymbol{k}}\left(n+\tfrac{1}{2}\right)\hbar\omega_c \, C_{\boldsymbol{k},n}^\dagger C_{\boldsymbol{k},n},
\end{equation}
and the interaction
\begin{equation}
    \begin{aligned}
        H_{\text{int}} & = \frac{1}{2A} \sum_{\substack{n_1,n_2,n_3,n_4 \\ \boldsymbol{k},\boldsymbol{k}',\boldsymbol{q} \in \text{BZ}}}\sum_{\boldsymbol{G}} V_{\boldsymbol{q}+\boldsymbol{G}} \, I_{[\boldsymbol{k}+\boldsymbol{q}],n_1; \boldsymbol{k},n_3}^{\boldsymbol{G}-\boldsymbol{G}_{\boldsymbol{k}+\boldsymbol{q}}} I_{[\boldsymbol{k}'-\boldsymbol{q}],n_2; \boldsymbol{k}',n_4}^{-\boldsymbol{G}-\boldsymbol{G}_{\boldsymbol{k}'-\boldsymbol{q}}} \\
        & \qquad \times C_{[\boldsymbol{k}+\boldsymbol{q}],n_1}^{\dagger} C_{[\boldsymbol{k}'-\boldsymbol{q}],n_2}^{\dagger} C_{\boldsymbol{k}',n_4} C_{\boldsymbol{k},n_3},
    \end{aligned}
\end{equation}
where $[\boldsymbol{q}] \equiv \boldsymbol{q} + \boldsymbol{G}_{\boldsymbol{q}}$ 
denotes the momentum folded back into the first Brillouin zone, and 
$\boldsymbol{G}_{\boldsymbol{q}}$ is the reciprocal lattice vector that 
performs this folding.

For Landau levels on a torus, the form factor admits the closed-form 
decomposition
\begin{equation}
    \mathcal{I}^{\boldsymbol{G}}_{\boldsymbol{k},n;\boldsymbol{p},n^\prime} 
    = \langle u_{\boldsymbol{k}}^{n\text{LL}}| e^{i \boldsymbol{G} \cdot \boldsymbol{r}}|u^{n^\prime\text{LL}}_{\boldsymbol{p}}\rangle 
    = \bar{\eta}_{\boldsymbol{G}} \, e^{\frac{i \ell^2}{2} \boldsymbol{G} \times \boldsymbol{p}} \langle u^{n\text{LL}}_{\boldsymbol{k}}|u^{n^\prime\text{LL}}_{\boldsymbol{p}-\boldsymbol{G}}\rangle,
\end{equation}
where $\bar{\eta}_{\boldsymbol{G}}$ is the parity factor of the reciprocal 
lattice vector $\boldsymbol{G}$, equal to $+1$ if $\boldsymbol{G}/2$ is 
itself a reciprocal lattice vector and $-1$ otherwise. The remaining 
intra-band overlap factor is given by the standard Landau-level result
\begin{equation}
    \langle u_{\boldsymbol{k}}^{n \mathrm{LL}} | u_{\boldsymbol{k}+\boldsymbol{q}}^{n^{\prime} \mathrm{LL}}\rangle 
    = \frac{(-1)^{n^{\prime}}}{\sqrt{n!\,n^{\prime}!}} \, \mathcal{L}_{n n^{\prime}}\!\left(\frac{\ell q^*}{\sqrt{2}}, \frac{\ell q}{\sqrt{2}}\right) e^{\ell^2\left(\frac{i}{2} \boldsymbol{k} \times \boldsymbol{q} - \frac{q^2}{4}\right)},
\end{equation}
with $q = q_x + i q_y$ the complex momentum, $q^* = q_x - i q_y$ its 
conjugate, and $\mathcal{L}_{n n^{\prime}}(x, y) = e^{x y} \partial_x^n \partial_y^{n^{\prime}} e^{-x y}$ 
the bivariate Laguerre-polynomial generator.

\begin{figure}[h]
    \centering
    \includegraphics[width=0.95\linewidth]{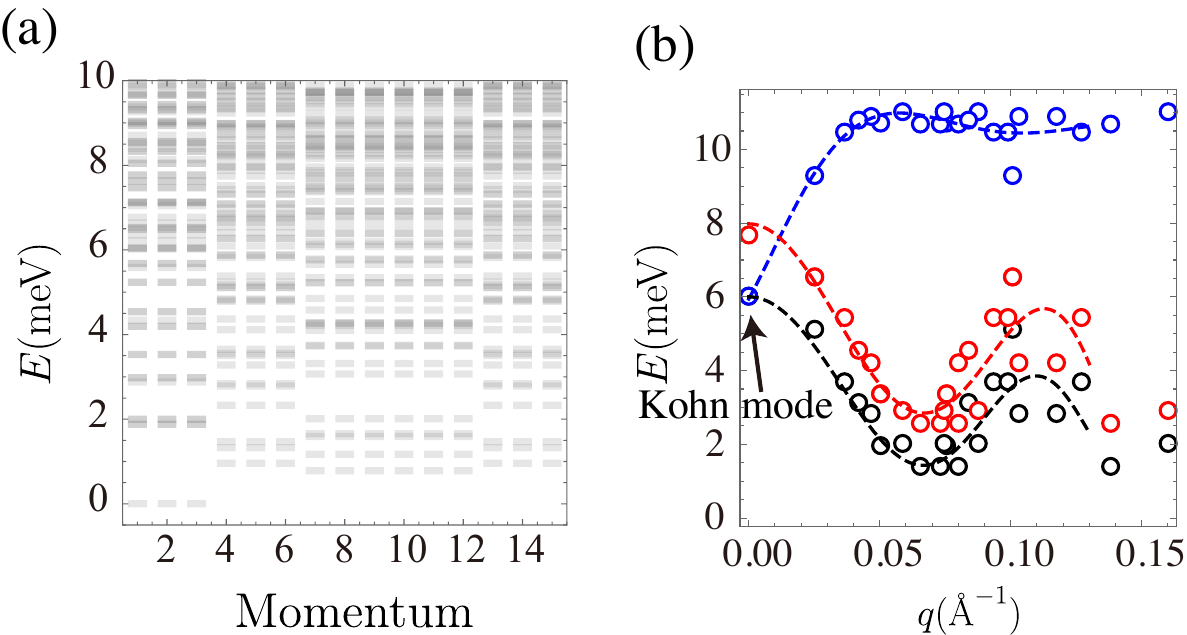}
    \caption{(a) Many-body spectrum of the two-Landau-level FQH model 
    on a $3\times 5$ cluster at $\nu = 1/3, \epsilon = 15$, showing the 
    threefold-degenerate Laughlin ground state. (b) Lowest interband 
    $|\Phi_{\text{inter}}\rangle$ (Kohn-like, blue), intraband 
    $|\Phi_{\text{intra}}\rangle$ (magneto-roton, red), and hybridization 
    mode $|\chi\rangle$ (black) dispersions obtained from ED-BSE. 
    At finite $\boldsymbol{q}$ the two modes hybridize through Coulomb 
    interactions; at $\boldsymbol{q} = 0$ the Kohn mode is decoupled 
    from the intra-LL sector by Galilean invariance, leading to the 
    abrupt level switching of $|\langle\Phi_{\text{inter}}|\chi\rangle|$ 
    in main-text Fig.~2(d). This long-wavelength decoupling is 
    qualitatively distinct from the smooth FCI hybridization observed 
    in main-text Fig.~2(c).}
    \label{fig:fqh}
\end{figure}

With the above Hamiltonian, we perform exact diagonalization on a 
$3\times 5$ momentum-space cluster at filling $\nu = 1/3$, retaining 
both the LLL and 1LL. The interaction strength is set to $\epsilon = 15$ 
to facilitate comparison with the FCI calculation of the main text 
(both correspond to the weak-mixing regime). The resulting many-body 
spectrum, shown in Fig.~\ref{fig:fqh}(a), exhibits the threefold-degenerate 
ground state characteristic of the $\nu = 1/3$ Laughlin state, well 
separated from the excited manifold by a clean gap.

We then apply the same ED-BSE variational procedure as in the main 
text—restricting either to the intraband sector ($|\Phi_{\text{intra}}\rangle$, 
particle-hole pairs within the LLL), to the interband sector 
($|\Phi_{\text{inter}}\rangle$, particle-hole pairs between LLL and 1LL), 
or to the full two-LL hybrid manifold ($|\chi\rangle$). 
The resulting dispersions, shown in Fig.~\ref{fig:fqh}(b), reveal two 
qualitatively distinct behaviors of the FQH system relative to the FCI:

At finite $\boldsymbol{q}$, the inter- and intra-LL modes 
do hybridize through the Coulomb interaction, as expected from the 
standard Landau-level mixing perturbation theory~\cite{PhysRevB.87.245425}. 
This is visible in Fig.~\ref{fig:fqh}(b) as a level repulsion between 
the bare interband (blue) and intraband (red) branches at intermediate 
$q$, with the lowest hybrid mode $|\chi\rangle$ (black) tracking the 
lower of the two.

At $\boldsymbol{q} = 0$, however, the inter-LL Kohn mode 
(arrow in Fig.~\ref{fig:fqh}(b)) is rigorously decoupled from the 
intra-LL sector by Galilean invariance: the center-of-mass cyclotron 
motion at $\boldsymbol{q} = 0$ is an exact eigenstate of the interacting 
Hamiltonian, and Coulomb interactions cannot mix with relative-motion 
excitations such as the magneto-roton. Consequently, the lowest hybrid 
state $|\chi\rangle$ at $\boldsymbol{q} = 0$ coincides with one of the 
two bare branches—whichever has lower energy—rather than acquiring an 
admixture of both. This produces the abrupt level switching of 
$|\langle\Phi_{\text{inter}}|\chi\rangle|$ across the gap-tuning curve 
in main-text Fig.~2(d).

This long-wavelength decoupling is the precise content of Kohn's 
theorem~\cite{PhysRev.123.1242} in the present context, and it is qualitatively 
distinct from the smooth FCI hybridization observed in main-text 
Fig.~2(c). It also implies that the optical response at $\boldsymbol{q} = 0$ 
in the FQH system is exhausted by the bare Kohn mode, with no spectral 
weight transferred to the magneto-roton sector—in sharp contrast to the 
double-peak optical signature of the FCI exciton-roton reported in 
main-text Fig.~3. The double-peak signature is therefore a uniquely 
lattice phenomenon, accessible only when the moir\'e potential breaks 
Galilean invariance and permits inter/intra-band hybridization at 
$\boldsymbol{q} = 0$.

% \section{Fractionalized quantized excitonic transition rate}
% In this section, we prove that the difference of the dichroic excitonic transition rate is quantized and equals the excitons' many-body Chern number.

% Starting from the Ferimi-Golden rules, the transition rate from the exciton to the another states in 

\end{document}